\documentclass
[preprint,prd,onecolumn,showpacs,showkeys,nobibnotes,nofootinbib,titlepage]{revtex4}%
\usepackage{amsfonts}
\usepackage{amsmath}
\usepackage{amssymb}
\usepackage{graphicx}
\usepackage{float}
\usepackage{rotating}%
\providecommand{\U}[1]{\protect\rule{.1in}{.1in}}

\ifx\pdfoutput\relax\let\pdfoutput=\undefined\fi
\newcount\msipdfoutput
\ifx\pdfoutput\undefined\else
\ifcase\pdfoutput\else
\ifx\paperwidth\undefined\else
\ifdim\paperheight=0pt\relax\else\pdfpageheight\paperheight\fi
\ifdim\paperwidth=0pt\relax\else\pdfpagewidth\paperwidth\fi
\fi\fi\fi
\begin{document}
\preprint{ }
\title[Kinematical Conformal Cosmology - Part I]{A Kinematical Approach to Conformal Cosmology}
\author{Gabriele U. Varieschi}
\affiliation{Department of Physics, Loyola Marymount University - Los Angeles, CA 90045,
USA\footnote{Email: gvarieschi@lmu.edu}}
\eid{ }
\author{}
\affiliation{}
\author{}
\affiliation{}

\begin{abstract}
We present an alternative cosmology based on conformal gravity, as originally
introduced by H. Weyl and recently revisited by P. Mannheim and D. Kazanas.
Unlike past similar attempts our approach is a purely kinematical application
of the conformal symmetry to the Universe, through a critical reanalysis of
fundamental astrophysical observations, such as the cosmological redshift and others.

As a result of this novel approach we obtain a closed-form expression for the
cosmic scale factor $R(t)$ and a revised interpretation of the space-time
coordinates usually employed in cosmology. New fundamental cosmological
parameters are introduced and evaluated. This emerging new cosmology does not
seem to possess any of the controversial features of the current standard
model, such as the presence of dark matter, dark energy or of a cosmological
constant, the existence of the horizon problem or of an inflationary phase.
Comparing our results with current conformal cosmologies in the literature, we
note that our kinematic cosmology is equivalent to conformal gravity with a
cosmological constant at late (or early) cosmological times.

The cosmic scale factor and the evolution of the Universe are described in
terms of several dimensionless quantities, among which a new cosmological
variable $\delta$ emerges as a natural cosmic time. The mathematical
connections between all these quantities are described in details and a
relationship is established with the original kinematic cosmology by L. Infeld
and A. Schild.

The mathematical foundations of our kinematical conformal cosmology will need
to be checked against current astrophysical experimental data, before this new
model can become a viable alternative to the standard theory.

\end{abstract}
\startpage{1}
\endpage{ }
\maketitle
\tableofcontents

\section{\label{sect:introduction}Introduction}

Modern cosmology has advanced very rapidly during these last decades,
producing an impressive model of the Universe, but our current understanding
is still troubled by many open questions and puzzles. Since the original
observations of cosmological redshift in spectral lines, done by V. M. Slipher
and E. P. Hubble almost one century ago and since the application of
Einstein's General Relativity to cosmological theoretical models, we have
progressed a long way towards our current picture, where the contents of the
Universe are described in terms of two main components, \textit{dark matter}
and \textit{dark energy}, accounting for most of the observed Universe, with
ordinary matter just playing a minor role.

The history of recent experimental observations which led to postulate the
existence of these two components is well known, as well as the many past and
current theoretical explanations (see for example \cite{Turner:1999md},
\cite{Freedman:2003ys}, \cite{Yao:2006px}, \cite{Perlmutter:1998np},
\cite{Riess:1998cb}, \cite{Riess:2004nr}, \cite{Riess:2006fw},
\cite{Spergel:2003cb}, \cite{Spergel:2006hy}), but since there is no evidence
of the real nature of dark matter and dark energy, we have to conclude that
our comprehension of the natural world is limited to only a very small
percentage of it (the ordinary matter component), a statement potentially very
embarrassing for cosmology and physics, if taken at face value.

Several alternatives to dark matter and dark energy have been proposed (for
comprehensive reviews see for example \cite{Mannheim:2005bf},
\cite{Schmidt:2006jt}, \cite{Nojiri:2006ri}, \cite{Clifton:2006jh},
\cite{Zakharov:2008zz} and references therein) which can be approximately
divided into two categories: those retaining the Newton-Einstein gravitational
paradigm, while introducing ad hoc corrections to explain dark matter and dark
energy and those breaking away substantially from established gravitational
theories. Following this second line of thought, we will concentrate our
attention on the theory of Conformal Gravity\ (CG), a fourth order extension
of Einstein's second order General Relativity (GR) as a possible framework for
the solution of current cosmological problems.

\section{\label{sect:conformal_gravity}Conformal Gravity}

\subsection{\label{sect:Weyl}Weyl's original proposal}

The idea of a possible \textquotedblleft conformal\textquotedblright%
\ generalization of Einstein's relativity was first developed by Hermann Weyl
in 1918 (\cite{Weyl:1918aa}, \cite{Weyl:1918ib}, \cite{Weyl:1919fi}). In his
pioneering work, Weyl introduced the so-called \textit{conformal} or
\textit{Weyl tensor}, a special combination of the Riemann tensor
$R_{\lambda\mu\nu\kappa}$, the Ricci tensor $R_{\mu\nu}=R^{\lambda}{}%
_{\mu\lambda\nu}$ and the curvature (or Ricci) scalar $R=R^{\mu}{}_{\mu}$ (see
\cite{Weinberg} p. 145):%

\begin{equation}
C_{\lambda\mu\nu\kappa}=R_{\lambda\mu\nu\kappa}-\frac{1}{2}(g_{\lambda\nu
}R_{\mu\kappa}-g_{\lambda\kappa}R_{\mu\nu}-g_{\mu\nu}R_{\lambda\kappa}%
+g_{\mu\kappa}R_{\lambda\nu})+\frac{1}{6}R\ (g_{\lambda\nu}g_{\mu\kappa
}-g_{\lambda\kappa}g_{\mu\nu}), \label{eqn2.1}%
\end{equation}
where, in particular, $C^{\lambda}{}_{\mu\lambda\nu}(x)$ is invariant under
the local transformation of the metric%

\begin{equation}
g_{\mu\nu}(x)\rightarrow\widehat{g}_{\mu\nu}(x)=e^{2\alpha(x)}g_{\mu\nu
}(x)=\Omega^{2}(x)g_{\mu\nu}(x). \label{eqn2.2}%
\end{equation}
The factor $\Omega(x)=e^{\alpha(x)}$ represents the amount of local
\textquotedblleft stretching\textquotedblright\ of the geometry, hence the
name \textquotedblleft conformal\textquotedblright\ for a theory invariant
under all possible local stretchings of the space-time.\footnote{The name
\textit{conformal}\ derives more precisely, \textquotedblleft from the
property that the transformation does not affect the angle between two
arbitrary curves crossing each other at some point, despite a local dilation:
the conformal group preserves angles\textquotedblright\ (quoted from
\cite{DiFrancesco:1997nk}).}

Weyl's ambitious original program was to introduce a new kind of geometry, in
relation to a unified theory of gravitation and electromagnetism where, in
addition to Eq. (\ref{eqn2.2}), the electromagnetic field would transform as
$A_{\mu}(x)\rightarrow\widehat{A}_{\mu}(x)=A_{\mu}(x)-e\ \partial_{\mu}%
\alpha(x)$. This theory was later abandoned with the advent of modern gauge
field interpretations of electrodynamics, retaining only terms such as
\textquotedblleft gauge transformation\textquotedblright\ or \textquotedblleft
gauge invariance,\textquotedblright\ which were introduced in reference to Eq.
(\ref{eqn2.2}) (for a brief history of conformal theories of gravitation from
1918 to 1988 see \cite{Schmidt:2006jt}, \cite{Schimming:2004yx}).

\subsection{\label{sect:fourth_order}Fourth order metric theories}

Following Weyl's idea, the conformally invariant generalizations of the
gravitational theory were found to be fourth order theories, as opposed to the
standard second order General Relativity. In other words, the field equations
originating from a conformally invariant Lagrangian contain derivatives up to
the fourth order of the metric with respect to the space-time coordinates.

Initially there was some ambiguity in the specific choice of the Lagrangian
and the related action for these new theories, but following work done by
Rudolf Bach \cite{Bach:1921}, Cornel Lanczos \cite{Lanczos:1938} and
others,\footnote{Even Albert Einstein used a conformally invariant formulation
in one of his papers in 1921 \cite{Einstein:1921}.} conformal gravity was
ultimately based on the conformal (or Weyl) action:\footnote{In this paper we
use a metric signature (-,+,+,+) and we follow the sign conventions of
Weinberg \cite{Weinberg}. We will use c.g.s. units when needed and all
fundamental constants, such as $c$ and $h$, will always be explicitly
introduced in every equation.}%

\begin{equation}
I_{W}=-\alpha_{g}\int d^{4}x\ (-g)^{1/2}\ C_{\lambda\mu\nu\kappa}%
\ C^{\lambda\mu\nu\kappa}, \label{eqn2.3}%
\end{equation}
or on the following equivalent expression (which differs from the previous one
by a topological invariant):%
\begin{equation}
I_{W}=-2\alpha_{g}\int d^{4}x\ (-g)^{1/2}\ \left(  R_{\mu\kappa}R^{\mu\kappa
}-\frac{1}{3}R^{2}\right)  , \label{eqn2.3.1}%
\end{equation}
where $g\equiv\det(g_{\mu\nu})$ and $\alpha_{g}$ is a gravitational coupling
constant (see \cite{Mannheim:2005bf}, \cite{Schimming:2004yx},
\cite{Mannheim:1988dj}, \cite{Kazanas:1988qa}).\footnote{In these cited
papers, $\alpha_{g}$ is referred to as a \textquotedblleft dimensionless
constant,\textquotedblright\ by working with natural units. Alternatively,
working with c.g.s. units, one can assign dimensions of an action to the
constant $\alpha_{g}$, so that the dimensionality of Eq. (\ref{eqn2.4}) will
also be correct.} Under the conformal transformation of Eq. (\ref{eqn2.2}),
the conformal tensor transforms as $C_{\lambda\mu\nu\kappa}\rightarrow
\widehat{C}_{\lambda\mu\nu\kappa}=e^{2\alpha(x)}C_{\lambda\mu\nu\kappa}%
=\Omega^{2}(x)C_{\lambda\mu\nu\kappa}$, while the Conformal Gravity action
$I_{W}$ above is completely locally conformal invariant and is actually the
unique general coordinate scalar action with such properties.

Variation of the Weyl action with respect to the metric led R. Bach
\cite{Bach:1921} to rewrite the gravitational field equation in the presence
of an energy-momentum tensor\footnote{We follow here the convention
\cite{Mannheim:2005bf}\ of introducing the energy-momentum tensor $T_{\mu\nu}$
so that the quantity $cT_{00}$ has the dimension of an energy density. For
example, we write the perfect fluid energy-momentum tensor as: $T_{\mu\nu
}=\frac{1}{c}[(\rho+p)U_{\mu}U_{\nu}+pg_{\mu\nu}]$.} $T_{\mu\nu}$ :%

\begin{equation}
W_{\mu\nu}=\frac{1}{4\alpha_{g}}\ T_{\mu\nu} \label{eqn2.4}%
\end{equation}
as opposed to the \textquotedblleft standard\textquotedblright\ Einstein's equation,%

\begin{equation}
R_{\mu\nu}-\frac{1}{2}g_{\mu\nu}\ R=-\frac{8\pi G}{c^{3}}\ T_{\mu\nu},
\label{eqn2.5}%
\end{equation}
where the \textquotedblleft Bach tensor\textquotedblright\ $W_{\mu\nu}$
\cite{Bach:1921}\ plays the role of the combination of the Ricci tensor and
curvature scalar on the left-hand side of Eq. (\ref{eqn2.5}). This tensor
$W_{\mu\nu}$ has a much more complex structure than those appearing in
Einstein's field equation. It is defined in a compact way as
\cite{Schmidt:1984bg}:%

\begin{equation}
W_{\mu\nu}=2C^{\alpha}{}_{\mu\nu}{}^{\beta}{}_{;\beta;\alpha}+C^{\alpha}%
{}_{\mu\nu}{}^{\beta}\ R_{\beta\alpha}, \label{eqn2.5.1}%
\end{equation}
but if one requests a form where the Weyl tensor does not explicitly appear,
the more complex structure for the Bach tensor will emerge
(\cite{Mannheim:1988dj}, \cite{Wood:2001ve}):%

\begin{align}
W_{\mu\nu}  &  =-\frac{1}{6}g_{\mu\nu}\ R^{;\lambda}{}_{;\lambda}+\frac{2}%
{3}R_{;\mu;\nu}+R_{\mu\nu}{}^{;\lambda}{}_{;\lambda}-R_{\mu}{}^{\lambda}%
{}_{;\nu;\lambda}-R_{\nu}{}^{\lambda}{}_{;\mu;\lambda}+\frac{2}{3}R\ R_{\mu
\nu}-2R_{\mu}{}^{\lambda}\ R_{\lambda\nu}+\label{eqn2.6}\\
&  +\frac{1}{2}g_{\mu\nu}\ R_{\lambda\rho}\ R^{\lambda\rho}-\frac{1}{6}%
g_{\mu\nu}\ R^{2},\nonumber
\end{align}
so that it involves derivatives up to the fourth order of the metric with
respect to space-time coordinates.

The mathematical complexity of the Bach tensor and of Eq. (\ref{eqn2.4}) was
one of the main reasons why the conformal theory of gravitation lost its
attractiveness, between the thirties and the sixties, while quantum field
theories were quickly progressing. A comprehensive review of the use of
conformal invariance in physics up to the 1960s can be found in Ref.
\cite{Fulton:1962bu}\ and references therein. Only in the seventies, it was
found that the fourth order theory is one-loop renormalizable
\cite{Stelle:1976gc},\ in contrast to standard general relativity, yielding a
revival of conformal gravity.

\subsection{\label{sect:Mannheim}Solutions to Conformal Gravity equations}

It was already known to Bach in 1921, that every static spherically symmetric
space-time, conformally related to the Schwarzschild-de Sitter solution, is a
static spherically symmetric solution of the Bach equation. In 1962, the
converse statement was shown by H. Buchdahl \cite{Buchdahl:1962}: every static
spherically symmetric solution of the Bach equation is conformally related to
the Schwarzschild-de Sitter solution (\cite{Reviewer:1}, \cite{Schmidt:1999vp}).

In this line of research, a solution of Bach's equation was published by P.
Mannheim and D. Kazanas (MK solution in the following) in 1989
(\cite{Mannheim:1988dj}, \cite{Kazanas:1988qa}) and also studied by R. Riegert
in his doctoral thesis \cite{Riegert:1986xw}. This was the exact and complete
exterior solution for a static, spherically symmetric source, in locally
conformal invariant Weyl gravity, i.e., the fourth order analogue of the
Schwarzschild exterior solution in General Relativity.

Solving Bach's Eq. (\ref{eqn2.4}), in the case $T_{\mu\nu}=0$, Mannheim and
Kazanas obtained a line element of the form
\begin{equation}
ds^{2}=-B(r)\ c^{2}dt^{2}+\frac{dr^{2}}{B(r)}+r^{2}d\psi^{2} \label{eqn2.7}%
\end{equation}
where $d\psi^{2}=d\theta^{2}+\sin^{2}\theta\ d\phi^{2}$ in spherical
coordinates and%

\begin{equation}
B(r)=1-\frac{\beta(2-3\beta\gamma)}{r}-3\beta\gamma+\gamma r-\kappa r^{2},
\label{eqn2.8}%
\end{equation}
with the parameters $\beta=\frac{MG}{c^{2}}\ (%
\operatorname{cm}%
)$, $\gamma\ (%
\operatorname{cm}%
^{-1})$, $\kappa\ (%
\operatorname{cm}%
^{-2})$ (again, we prefer to show explicitly constants such as the speed of
light $c$ in all formulas), where $M$ is the mass of the (spherically
symmetric) source. The familiar Schwarzschild solution is recovered in the
limit for $\gamma,\kappa\rightarrow0$, in the equations above. The other two
parameters are interpreted by MK \cite{Mannheim:1988dj} in the following way:
$\kappa$ and the corresponding term $\kappa r^{2}$ should indicate a
background De Sitter space-time which would be important only at cosmological
distances, since $\kappa$ should have a very small value. On the other hand,
$\gamma$ measures the departure from the Schwarzschild metric, with the
$\gamma r$ term becoming significant over galactic distance scales.

In other words, for values of $\gamma\thickapprox10^{-28}-10^{-30}\
\operatorname{cm}%
^{-1}$, which is about the value of the inverse Hubble length, the standard
Newtonian $\frac{1}{r}$ term still dominates at smaller distances, so that
this theory would yield the same experimental success of General Relativity at
the scale of the solar system. The three classic tests of GR, namely the
gravitational redshift, the gravitational bending of light and the precession
of planetary orbits, would still hold for conformal gravity at the solar
system scale \cite{Mannheim:2005bf}. The only additional test of the
gravitational theory, at this distance scale, which has not been analyzed yet
in the MK theory, is the well-known decay of the orbit of a binary pulsar
(\cite{Hulse:1974eb}, \cite{Taylor:1989sw}, \cite{Weisberg:2002nv},
\cite{Weisberg:2004hi}).

Considering larger galactic distances, the contribution of the additional
$\gamma r$ term might explain the flat galactic rotation curves, without the
need of dark matter. This important connection to the dark matter problem and
the galactic rotation curves was subsequently studied in great detail by
Mannheim in a series of papers (\cite{Mannheim:1992vj}, \cite{Mannheim:1993rs}%
, \cite{Mannheim:1993pi}, \cite{Mannheim:1995ti}, \cite{Mannheim:1995eb},
\cite{Mannheim:1996jt}, \cite{Mannheim:1996rv}, \cite{Mannheim:2005bf}),
showing that it is possible to fit the experimental galactic rotation data
with theoretical curves based on conformal gravity, with the same level of
accuracy of current dark matter theories (see Fig. 1 of Ref.
\cite{Mannheim:1996rv} or Ref. \cite{Mannheim:2005bf}, for example), thus
establishing conformal gravity as a viable alternative to the dark matter hypothesis.

When we apply conformal gravity to a galaxy, we need to specify in more
details the role of the parameters $\beta$ and $\gamma$. Again, Mannheim has
shown that the Newtonian $\frac{1}{r}$ potential can be recovered for short
distances, as a solution of a fourth order Poisson equation for the
gravitational potential $\phi$, as opposed to the standard second order
equation (see \cite{Mannheim:2005bf}, Sect. 4.2 for details). The resulting
exterior potential for a single star source is of the form:%

\begin{equation}
\phi^{\ast}(r>R)=-\frac{\beta^{\ast}c^{2}}{r}+\frac{\gamma^{\ast}c^{2}r}{2}
\label{eqn2.9}%
\end{equation}
where $\beta^{\ast}$ and $\gamma^{\ast}$ are the individual parameters for a
system composed of a single star (i.e., $\beta^{\ast}=\frac{M_{\odot}G}{c^{2}%
}$, where we use the solar mass $M_{\odot}$ as a reference mass for a stellar
object). In first approximation, for a system of $N^{\ast}$ stars in a galaxy,
we would expect to introduce overall $\beta$ and $\gamma$ parameters which are
linear in the number of sources: $\beta=N^{\ast}\beta^{\ast}\ ;\ \gamma
=N^{\ast}\gamma^{\ast}$.

A more detailed analysis was done by Mannheim (\cite{Mannheim:1996jt}%
,\cite{Mannheim:1996rv}) on a representative sample of eleven spiral galaxies,
fitting their rotational velocity curves using the conformal gravity approach
(Fig. 1 of Ref. \cite{Mannheim:1996rv} or Ref. \cite{Mannheim:2005bf}
illustrates this detailed fitting). The galaxies were modeled with a
thin/thick disk potential with the addition of a spherical central bulge
region if necessary. The luminous Newtonian contribution was found to account
well for the initial rise of the rotation curve from the center of the galaxy
($r=0)$ up to a peak at $r=2.2\ r_{0}$, where $r_{0}$ is the scale length of
the galaxy and $r$ is the radial coordinate. The centripetal acceleration due
to just the luminous matter distribution would yield the standard Keplerian
term $\frac{v_{lum}^{2}}{r}\rightarrow\frac{N^{\ast}\beta^{\ast}c^{2}}{r^{2}}%
$, outside the optical disk. The number of stars $N^{\ast}$ in each galaxy was
computed by fitting the rotational curve, just due to the luminous Newtonian
contribution, to the experimental value at the peak for $r=2.2\ r_{0}$.

The discrepancy observed between the experimental data and the Keplerian
prediction, for distances larger than the peak distance, was then modeled with
parameters from conformal gravity. In particular, the last experimentally
observed value for the rotational acceleration $\frac{v_{last}^{2}}{r}$\ of
the sample galaxies, was found to be well explained by a two parameter formula
(in addition to the standard Keplerian term introduced above):%

\begin{equation}
\frac{v_{last}^{2}}{r}=\frac{N^{\ast}\beta^{\ast}c^{2}}{r^{2}}+\frac{N^{\ast
}\gamma^{\ast}c^{2}}{2}+\frac{\gamma_{0}c^{2}}{2}. \label{eqn2.10}%
\end{equation}

In the previous equation, the first term on the right-hand side is the
standard Keplerian one, while the two additional terms come from the conformal
theory, without any need of dark matter contributions. The two additional
universal parameters are evaluated from the detailed fitting of the
experimental curves as follows \cite{Mannheim:1996rv}:%

\begin{equation}
\gamma^{\ast}=5.42\times10^{-41}%
\operatorname{cm}%
^{-1};\ \gamma_{0}=3.06\times10^{-30}%
\operatorname{cm}%
^{-1} \label{eqn2.11}%
\end{equation}
and their interpretation is analogous to the $\gamma$ parameter of the MK
solution of Eq. (\ref{eqn2.8}).

The presence of two gamma parameters is also explained by Mannheim: the
$\gamma^{\ast}N^{\ast}$ term is the gamma parameter of the specific galaxy
being analyzed, being the product of the single star contribution
$\gamma^{\ast}$ times the number of stars $N^{\ast}$ in the galaxy being
considered. The more universal $\gamma_{0}\simeq3.06\times10^{-30}%
\operatorname{cm}%
^{-1}$ represents a cosmological gamma parameter, presumably due to the
combined effect of all the galaxies (see discussion on page 416 of
\cite{Mannheim:2005bf}). This term would affect the space-time geometry even
in regions far away from matter sources, introducing an \textquotedblleft
universal acceleration\textquotedblright\ $\frac{\gamma_{0}c^{2}}%
{2}=1.38\times10^{-9}\ \frac{%
\operatorname{cm}%
}{%
\operatorname{s}%
^{2}}$ which is close to similar universal acceleration parameters, such as
those introduced by the Modified Newtonian Dynamics (MOND) theory by M.
Milgrom and others (\cite{Milgrom:1983ca}, \cite{Milgrom:1983pn},
\cite{Bekenstein:1984tv}, \cite{Bekenstein:2004ne}).

In view of the very successful fitting of the experimental galactic rotation
curves, shown in \cite{Mannheim:1996rv}, we will consider here the Conformal
Gravity model as a viable alternative to the dark matter hypothesis. In
particular, we will retain the cosmological parameter $\gamma_{0}$, which will
be used in our subsequent analysis, but we will need to reconsider its meaning
and value later in this work.

\section{\label{sect:conformal_cosmology}Conformal Cosmology}

As outlined in the previous section, we will assume that Conformal Gravity is
a possible alternative gravitational theory, therefore the next logical step
is to construct a cosmology based on these new ideas. In fact, many conformal
cosmologies exist in the literature, including the one proposed by Mannheim in
another series of papers (\cite{Mannheim:1989jh}, \cite{Mannheim:1991cz},
\cite{Mannheim:1991ez}, \cite{Mannheim:1996cd}, \cite{Mannheim:1998ew},
\cite{Mannheim:1999nc}, \cite{Mannheim:2001kk}, \cite{Mannheim:2003xy},
\cite{Mannheim:2007ug}, \cite{Mannheim:2008sq}). Mannheim's cosmology is based
on the construction of a traceless (as required by the conformal theory)
energy-momentum tensor $T_{\mu\nu}$, in a theory in which the action is built
out of fields rather than particles, using a spontaneous symmetry breaking
mechanism in order to obtain particle masses. This modern approach elegantly
overcomes the original objection to a conformal, scaleless theory, which would
strictly require all particles to be massless, but is not free from
theoretical controversy (\cite{Mannheim:2007ug}, \cite{Flanagan:2006ra}).

Other \textquotedblleft conformal\textquotedblright\ cosmologies exist in the
literature (see for example \cite{Faraoni:1998qx}, \cite{Behnke:2001nw},
\cite{Schmidt:2006jt}), based on similar approaches, but none of these has
become a popular cosmological alternative to the standard model or even to
cosmologies based on the MOND approach, including its latest relativistic
version (Tensor-Vector-Scalar gravity, TeVeS, \cite{Bekenstein:2004ne}). In
our opinion, all these conformal cosmologies do not fully explain the
connection between the assumed conformal symmetry and the physical reality of
our Universe, as determined by cosmological observations. Therefore, we seek
here an alternative approach, which doesn't require the field theory
formalism, but is based on a critical analysis of the foundations of
observational cosmology, starting with cosmological redshift.

\subsection{\label{sect:from_SSC_to_RW}From Static Standard Coordinates to the
Robertson-Walker Metric}

To introduce the discussion of cosmological redshift, it is necessary to
analyze here in more details the transformation of the coordinates related to
the MK solution, in particular the transformation from Static Standard
Coordinates (SSC) to the Robertson-Walker (RW) metric. This is another
fundamental aspect of Conformal Gravity: the CG solution is able to
interpolate smoothly between the static Schwarzschild solution and the
classical Robertson-Walker metric. We will follow again Mannheim and Kazanas
(\cite{Mannheim:2005bf}, \cite{Mannheim:1988dj}, \cite{Kazanas:1988qa}), but
we will use a slightly different notation and interpretation, for the
different sets of coordinates used in the following. Another complete
description of the necessary coordinate and conformal transformations from the
Schwarzschild-de Sitter solution to the Mannheim-Kazanas solution can be found
in Ref. \cite{Schmidt:1999vp}.

We start again from the line element given by Eqs. (\ref{eqn2.7}) and
(\ref{eqn2.8}), but we consider now regions far away from matter
distributions, thus ignoring the matter dependent $\beta$ term. In view of the
discussion in the previous section, we could identify the $\gamma$ parameter
in Eq. (\ref{eqn2.8}) with $\gamma_{0}\simeq3.06\times10^{-30}%
\operatorname{cm}%
^{-1}$, as in Eq. (\ref{eqn2.11}). However, this value refers to a sample of
eleven galaxies, where the rotational motion data being fitted by the
conformal gravity theory cover a range of distances of a few kiloparsec, from
the center of each galaxy.

In our next paper \cite{Varieschi:2008va} we will argue that parameters such
as $\gamma$ are better determined by \textquotedblleft local\textquotedblright%
\ measurements on a short distance scale and not on the kiloparsec scale.
Mannheim's value of $\gamma_{0}\simeq3.06\times10^{-30}%
\operatorname{cm}%
^{-1}$ can therefore provide a useful order of magnitude for this quantity,
but we will determine later its \textquotedblleft current\textquotedblright%
\ value from more local measurements.

We will use the greek letter $\kappa$ for the additional integration constant
in the MK solution, instead of $k$ used in the original references. In
particular, we retain here the $\kappa r^{2}$ \textquotedblleft cosmological
background\textquotedblright\ term that was dropped by Mannheim in his latest
analysis \cite{Mannheim:2005bf}, which on the contrary will play an essential
role in our cosmology. We therefore write $B(r)$ as:%

\begin{equation}
B(r)=1+\gamma r-\kappa r^{2} \label{eqn3.1}%
\end{equation}
so that the line element becomes:%

\begin{equation}
ds^{2}=-\left(  1+\gamma r-\kappa r^{2}\right)  \ c^{2}dt^{2}+\frac{dr^{2}%
}{\left(  1+\gamma r-\kappa r^{2}\right)  }+r^{2}d\psi^{2}, \label{eqn3.2}%
\end{equation}
in what we will call the Static Standard Coordinates - SSC $(r,t,\theta,\phi)$
in the following. These are the coordinates we use to carry out all our
standard laboratory measurements, with our current units of length, time, mass
and others.

With a first coordinate transformation:\footnote{The angular coordinates
$\theta$ and $\phi$, as well as the quantity $d\psi^{2}=d\theta^{2}+\sin
^{2}\theta\ d\phi^{2}$, are not changed by any of the transformations
performed in this section. Therefore, we will not rename these angular
coordinates. Note also the inverse transformation: $r=\frac{\rho}%
{(1-\gamma\rho/4)^{2}+\kappa\rho^{2}/4};\ t=\int\frac{d\tau}{R(\tau)}$.}%

\begin{align}
\rho &  =\frac{4r}{2\sqrt{1+\gamma r-\kappa r^{2}}+2+\gamma r}\label{eqn3.3}\\
\tau &  =\int R(t)\ dt\nonumber
\end{align}
the metric, as a line element, becomes (\cite{Mannheim:1988dj},
\cite{Kazanas:1988qa}, \cite{Mannheim:2005bf}):%

\begin{equation}
ds^{2}=\frac{1}{R^{2}(\tau)}\ \frac{[1-\rho^{2}(\frac{\gamma^{2}}{16}%
+\frac{\kappa}{4})]^{2}}{[(1-\frac{\gamma\rho}{4})^{2}+\frac{\kappa\rho^{2}%
}{4}]^{2}}\left\{  -c^{2}d\tau^{2}+\frac{R^{2}(\tau)}{[1-\rho^{2}(\frac
{\gamma^{2}}{16}+\frac{\kappa}{4})]^{2}}(d\rho^{2}+\rho^{2}d\psi^{2})\right\}
. \label{eqn3.4}%
\end{equation}
At this point it is convenient to redefine the combination of parameters
$\gamma$ and $\kappa$, in Eq. (\ref{eqn3.4}), as follows:%

\begin{equation}
\frac{\gamma^{2}}{16}+\frac{\kappa}{4}=-\frac{k}{4}, \label{eqn3.5}%
\end{equation}
where $k$ will be ultimately linked to the \textquotedblleft trichotomy
constant\textquotedblright\ of a Robertson-Walker (RW) metric. Equation
(\ref{eqn3.4}) can be rewritten as:%

\begin{equation}
ds^{2}=\frac{1}{R^{2}(\tau)}\ \frac{[1+\frac{k}{4}\rho^{2}]^{2}}%
{[1-\frac{\gamma}{2}\rho-\frac{k}{4}\rho^{2}]^{2}}\left\{  -c^{2}d\tau
^{2}+\frac{R^{2}(\tau)}{[1+\frac{k}{4}\rho^{2}]^{2}}(d\rho^{2}+\rho^{2}%
d\psi^{2})\right\}  . \label{eqn3.6}%
\end{equation}

As noted by Mannheim and Kazanas \cite{Mannheim:1988dj}, the metric above is
conformal to a RW metric in isotropic form. All we need is to apply a
conformal transformation, such as the one in Eq. (\ref{eqn2.2}), to the metric
tensor $g_{\mu\nu}(\rho,\tau)$ defined through Eq. (\ref{eqn3.6}), to obtain a
new metric $\widehat{g}_{\mu\nu}(\rho,\tau)$ in the RW isotropic form.
Precisely, we will \textquotedblleft stretch\textquotedblright\ the space-time
fabric, multiplying the last equation by the factor%

\begin{equation}
\Omega^{2}(\rho,\tau)=R^{2}(\tau)\ \frac{[1-\frac{\gamma}{2}\rho-\frac{k}%
{4}\rho^{2}]^{2}}{[1+\frac{k}{4}\rho^{2}]^{2}}, \label{eqn3.7}%
\end{equation}
which depends on the space-time coordinates. We will then replace the metric
as follows:%

\begin{equation}
g_{\mu\nu}(\rho,\tau)\rightarrow\widehat{g}_{\mu\nu}(\rho,\tau)=\Omega
^{2}(\rho,\tau)\ g_{\mu\nu}(\rho,\tau)=R^{2}(\tau)\ \frac{[1-\frac{\gamma}%
{2}\rho-\frac{k}{4}\rho^{2}]^{2}}{[1+\frac{k}{4}\rho^{2}]^{2}}\ g_{\mu\nu
}(\rho,\tau). \label{eqn3.8}%
\end{equation}
Therefore we obtain:%

\begin{equation}
d\widehat{s}^{2}=-c^{2}d\tau^{2}+\frac{R^{2}(\tau)}{[1+\frac{k}{4}\rho
^{2}]^{2}}(d\rho^{2}+\rho^{2}d\psi^{2}), \label{eqn3.9}%
\end{equation}
and the metric is now in the form known as the \textquotedblleft
isotropic\textquotedblright\ Robertson-Walker. In the previous equations we
did not use different symbols for the coordinates after the conformal
transformation, but we kept the previous set of $\rho,\tau$ coordinate. The
theory of local conformal transformations of the metric indicates that we can
always choose the new coordinates of a point, after the local stretching, so
that they correspond to the old coordinates $\rho,\tau$ of the original point
before the stretching (see \cite{Fulton:1962bu} for a detailed discussion of
conformal transformations in physics).

The above transformation implies a change of the line element itself, which is
stretched by the same amount%

\begin{equation}
d\widehat{s}^{2}=\Omega^{2}(\rho,\tau)\ ds^{2} \label{eqn3.9.1}%
\end{equation}
and this \textquotedblleft gauge transformation\textquotedblright\ will
ultimately result in a redefinition of the local measuring rods and clocks,
which will be a key feature of our cosmology. Another coordinate
transformation will lead from the isotropic form of RW metric to the standard
RW metric:\footnote{The inverse transformation of Eq. (\ref{eqn3.11}) is:
$\rho=2(\frac{1-\sqrt{1-k\rho^{\prime2}}}{k\rho^{\prime}});\ \tau=\tau
^{\prime}$, where the minus sign in front of the square root selects the
correct branch of the graph of the function considered. For $k=0$ it reduces
simply to $\rho=\rho^{\prime}$.}%

\begin{align}
\rho^{\prime}  &  =\frac{\rho}{1+\frac{k}{4}\rho^{2}}\label{eqn3.11}\\
\tau^{\prime}  &  =\tau\nonumber
\end{align}
and the metric becomes%

\begin{equation}
d\widehat{s}^{2}=-c^{2}d\tau^{\prime2}+R^{2}(\tau^{\prime})\left[  \frac
{d\rho^{\prime2}}{1-k\rho^{\prime2}}+\rho^{\prime2}d\psi^{2}\right]  .
\label{eqn3.12}%
\end{equation}
In this expression the parameter $k$ is still linked to $\gamma$ and $\kappa$,
through Eq. (\ref{eqn3.5}), or equivalently:%

\begin{equation}
k=-\frac{\gamma^{2}}{4}-\kappa. \label{eqn3.13}%
\end{equation}

It is customary for the so-called trichotomy constant of a Robertson-Walker
(RW) metric to have values $0,\pm1$. This can be accomplished with a final
rescaling of the coordinates, of the constant $k$ and of the scale factor $R$,
as follows:%

\begin{align}
\mathbf{k}  &  =\frac{k}{\left\vert k\right\vert }=0,\pm1\label{eqn3.14}\\
\mathbf{r}  &  =\sqrt{\left\vert k\right\vert }\rho^{\prime}\nonumber\\
\mathbf{t}  &  =\tau^{\prime}\nonumber\\
\mathbf{R}(\mathbf{t})  &  =\frac{R(\tau^{\prime})}{\sqrt{\left\vert
k\right\vert }},\nonumber
\end{align}
where we use bold symbols $\mathbf{k,r,t,R}$ to denote quantities after this
last transformation.\footnote{In the special case $k=0$ the transformation in
Eq. (\ref{eqn3.14}) should actually read: $\mathbf{k}=0$; $\mathbf{r}%
=\rho^{\prime}$; $\mathbf{t}=\tau^{\prime}$ and $\mathbf{R(t)}=R(\tau^{\prime
})$.} We can finally obtain the standard Robertson-Walker form of the
metric:\footnote{We note that, due to the transformations of Eq.
(\ref{eqn3.14}), the quantities $\mathbf{r}$ and $\mathbf{k}$ are now
dimensionless, while the factor $\mathbf{R}$ acquires the dimension of length.
We will not follow the common alternative normalization, with a dimensionless
scale factor, which is sometimes found in the literature.}%

\begin{equation}
d\widehat{s}^{2}=-c^{2}d\mathbf{t}^{2}+\mathbf{R}^{2}(\mathbf{t})\left[
\frac{d\mathbf{r}^{2}}{1-\mathbf{kr}^{2}}+\mathbf{r}^{2}d\psi^{2}\right]
;\ \mathbf{k}=0,\pm1. \label{eqn3.15}%
\end{equation}

We recall that the RW metric in the previous equation can be expressed
equivalently in the so-called curvature normalized form:%

\begin{align}
d\widehat{s}^{2}  &  =-c^{2}d\mathbf{t}^{2}+\mathbf{R}^{2}(\mathbf{t})\left[
d\mathbf{\chi}^{2}+S_{\mathbf{k}}^{2}(\mathbf{\chi})d\psi^{2}\right]
\label{eqn3.15.1}\\
S_{\mathbf{k}}(\mathbf{\chi})  &  \mathbf{\equiv}%
\begin{Bmatrix}
\sin\chi\ ; & \mathbf{k}=+1\\
\chi\ ; & \mathbf{k}=0\\
\sinh\chi\ ; & \mathbf{k}=-1
\end{Bmatrix}
,\nonumber
\end{align}
for closed, flat or open universes respectively. The \textit{comoving
coordinate} $\mathbf{\chi}$ is also dimensionless and the connection with the
$\mathbf{r}$ coordinate in Eq. (\ref{eqn3.15}) is due to the simple relation:%

\begin{equation}
\int_{\mathbf{0}}^{\mathbf{r}}\frac{d\mathbf{r}^{\prime}}{\sqrt{1-\mathbf{kr}%
^{\prime2}}}=%
\begin{Bmatrix}
\arcsin\mathbf{r\ ;} & \mathbf{k}=+1\\
\mathbf{r\ ;} & \mathbf{k}=0\\
\operatorname{arcsinh}\mathbf{r\ ;} & \mathbf{k}=-1
\end{Bmatrix}
=S_{\mathbf{k}}^{-1}(\mathbf{r})=\chi. \label{eqn3.15.2}%
\end{equation}

Another important quantity for our discussion is the \textit{conformal time}
interval $d\eta$, usually defined as an interval $c\ d\mathbf{t}$ divided by
the scale factor $\mathbf{R}(\mathbf{t})$:%

\begin{equation}
d\eta=\frac{cd\mathbf{t}}{\mathbf{R(t)}}=\sqrt{\left\vert k\right\vert }%
\frac{cd\mathbf{t}}{R\mathbf{(}t\mathbf{)}}=\sqrt{\left\vert k\right\vert }cdt
\label{eqn3.15.3}%
\end{equation}
which is essentially equivalent to the SSC time interval $dt$, in view of Eqs.
(\ref{eqn3.3}), (\ref{eqn3.11}), (\ref{eqn3.14}) and was in fact already
introduced by the transformations of Eq. (\ref{eqn3.3}). Using the RW metric
in the form of Eq. (\ref{eqn3.15.1}) we obtain a well-known and simple
expression for the null geodesic $d\widehat{s}^{2}=0$, corresponding to the
propagation of a light signal in the radial direction ($d\psi=0$):%

\begin{equation}
d\chi=\frac{d\mathbf{r}}{\sqrt{1-\mathbf{kr}^{2}}}=-\frac{cd\mathbf{t}%
}{\mathbf{R(t)}}=-\sqrt{\left\vert k\right\vert }\ cdt=-d\eta,
\label{eqn3.15.4}%
\end{equation}
thus establishing a direct connection between the comoving coordinate $\chi$,
the conformal time $\eta$ and the SSC time coordinate $t$.\footnote{For $k=0$
we recall that $\mathbf{R(t)}=R(t)$, therefore Eq. (\ref{eqn3.15.4}) should be
written as $d\chi=-cdt=-d\eta$, omitting the $\sqrt{\left\vert k\right\vert }$
factor.} We note that the second equality in Eq. (\ref{eqn3.15.4}) is only
valid for a null geodesic, i.e., $\chi$ and $\eta$ are simply related to each
other only when describing the propagation of a light signal. In this case the
minus signs in the previous equation indicate that we are following a light
ray propagating in the negative\ radial direction ($d\chi<0$) for increasing
conformal time ($d\eta>0$).

Summarizing this section: the coordinate transformations described above
allowed us to connect the original Static Standard Coordinates $(r,t,\theta
,\phi)$, used by Conformal Gravity to solve the problem of the rotational
galactic curves without resorting to dark matter, to the cosmological comoving
coordinates $(\mathbf{r},\mathbf{t},\theta,\phi)$, commonly used together with
Eq. (\ref{eqn3.15}) as the basis of standard cosmology. We will continue to
use normal and bold characters in the following to differentiate between these
two sets of coordinates.

\subsection{\label{sect:new_red_shift}An alternative interpretation of the
cosmological redshift}

One of the foundations of observational cosmology is the well known
cosmological redshift of galaxies, which is usually related to the expansion
of the Universe. It is customary (see \cite{Weinberg}, \cite{Peebles:1994xt},
\cite{Kolb:1990vq}, \cite{Lang:1999ry}, \cite{Peacock:1999ye}, or any other
General Relativity - Cosmology textbook) to consider light emitted by a
distant galaxy at (comoving) coordinates $(\mathbf{r,t},\theta,\phi)$ and
reaching us at the origin of the coordinates $\mathbf{r}=0$ and at time
$\mathbf{t}_{0}$ (present time). The time of emission $\mathbf{t}$ is
therefore in the past, i.e., $\mathbf{t<t}_{0}$, or $\mathbf{t}_{0}%
\mathbf{-t}>0$ is the \textquotedblleft look-back\textquotedblright%
\ time.\footnote{In this way, integrating Eq. (\ref{eqn3.15.4}) for light
emitted at coordinate $\chi$, at conformal time $\eta$, and reaching us at the
origin ($\chi=0$) at our present conformal time $\eta_{0}$, we obtain:
$\chi=\sqrt{\left\vert k\right\vert }\ c(t_{0}-t)=\eta_{0}-\eta$.} The
redshift parameter $z$ is related to the cosmic scale factor $\mathbf{R(t)}$,
or to the change in the radiation wavelength/frequency, through the standard expression:%

\begin{equation}
1+z=\frac{\mathbf{R(t}_{0}\mathbf{)}}{\mathbf{R(t)}}=\frac{\lambda_{0}%
}{\lambda}=\frac{\nu}{\nu_{0}}, \label{eqn3.16}%
\end{equation}
where, quoting from Weinberg (see \cite{Weinberg}, pages 416-417):
\textquotedblleft... $\nu$ and $\lambda$ are the frequency and wavelength of
the light if observed near the place and time of emission, and hence
presumably take the values measured when the same atomic transition occurs in
terrestrial laboratories, while $\nu_{0}$ and $\lambda_{0}$ are the frequency
and wavelength of the light observed after its long journey to
us.\textquotedblright

Given this standard view of the redshift, it has always been considered a
serious misconception to interpret the expansion of the Universe as if,
\textquotedblleft space itself is swelling up,\textquotedblright\ thus causing
galaxies to separate. Numerous textbooks are quick to point out this
potentially erroneous interpretation (see for example \cite{Peacock:1999ye},
\cite{Webb:1999}), explaining that galaxies separate, \textquotedblleft like
coins glued on an inflating balloon,\textquotedblright\ without altering their
intrinsic dimensions, or that two massless objects set up at rest with respect
to each other will show no tendency to separate, due to cosmological expansion.

However, an analysis of the literature of cosmological theories also reveals
that other possible interpretations of the redshift, apart from the standard
general relativistic expansion, were considered. Many alternative theories
exist such as the \textit{kinematic cosmology}\ by Infeld and Schild
(\cite{1945Nature}, \cite{1946PhDT.........6S}, \cite{PhysRev.68.250},
\cite{PhysRev.70.410}), which is also based on the conformal gauge
transformation of Eq. (\ref{eqn3.9.1}) as well as the cosmological principle
and the constancy of the speed of light. In this theory all possible
cosmological models based on these assumptions are analyzed and classified,
leading to different possible interpretations of the redshift, ranging from
standard Doppler effect to the purely \textquotedblleft gravitational
redshift\textquotedblright\ effect that we will also employ in our model. We
will later compare our results to the different models proposed by Infeld and Schild.

These conformally-flat-spacetime models were recently also studied by others
(\cite{1994ApJ...434..397E}, \cite{1997ApJ...479...40E},
\cite{1998ApJ...508..129Q}) and were also considered in other theories such as
the Hoyle-Narlikar cosmology \cite{Narlikar:1986ce}. This model, for example,
assumes a non standard interpretation of the cosmological redshift, i.e.,
since the atomic radiation wavelength is inversely proportional in first
approximation to the mass of the electron involved in the atomic transition,
the ratio $\lambda_{0}/\lambda$ is simply assumed to correspond to the value
of the (variable) electron mass at different epochs: $\lambda_{0}%
/\lambda=m_{e}(\mathbf{t})/m_{e}(\mathbf{t}_{0})$. Hoyle-Narlikar then
implemented their model, assuming a conformally invariant theory where masses
scale as $\widehat{m}=\Omega^{-1}m$, adding a variable gravitational constant
$G$, whose variation is based on a large numbers hypothesis, similar to the
original Dirac argument (\cite{Dirac:1937ti}, \cite{Dirac:1938mt}) and finally
proposed mechanisms of particle creation, in line with previous steady-state cosmologies.

While we do not agree with such theories, we share the idea that the redshift
ratio $\lambda_{0}/\lambda$ might be disclosing to us information about the
emission/absorption process at different cosmological epochs. In this line of
reasoning, we recall that modern metrology (see metrology web-sites
\cite{NIST}, \cite{BIPM} and references therein) defines our basic units of
length and time using non-gravitational physics, through a reference atomic
wavelength or frequency, so that our \textit{meter}\footnote{The meter was
recently redefined as the length of the path travelled by light in vacuum
during a time interval of $1/299\ 792\ 458$ of a second. This definition
assumes an (exact) speed of light in vacuum: $c=299\ 792\ 458\ m\ s^{-1}$. In
this way the unit of length is basically defined through the unit of time,
therefore not altering the validity of our discussion.} is just some multiple
$n_{m}$ of an atomic reference wavelength $\lambda_{m}$, or equivalently the
\textit{second} is a multiple $n_{s}$ of the inverse of some atomic reference
frequency $\nu_{s}$:%
\begin{align}
1\ meter  &  \equiv n_{m}\ \lambda_{m}\label{eqn3.17}\\
1\ second  &  \equiv n_{s}\ \frac{1}{\nu_{s}}.\nonumber
\end{align}

Since our space-time units ultimately have an atomic definition based on
emission/absorption of radiation, a possible \textquotedblleft
swelling\textquotedblright\ or dilation of the space-time fabric at any level,
from the atomic to the galactic scale, could never be detected using currently
defined meter sticks and clocks, because these would be \textquotedblleft
dilated\textquotedblright\ by the same amount.

In other words, it is only possible to base our space-time units on the
\textit{current }and\textit{ local} values of wavelength or frequency of some
standard reference atomic transition, but we cannot be absolutely certain that
these reference wavelengths or frequencies are invariable and constant
throughout the Universe and at all cosmological times. A possible variation of
these reference wavelengths and frequencies would be also related to the
well-known problem of the time variation of the \textit{universal constants}
(for modern reviews see \cite{Barrow1}, \cite{Uzan:2002vq}, \cite{Okun:2003wc}).

The logical connection between a possible conformal symmetry of the Universe,
dealing with stretchings and dilations of the metric, and the previous
discussion of changes and variations in our meter sticks and clock rates,
should induce a revision of the redshift mechanism. In particular, the
observed galactic redshift might be interpreted, in part or completely, as due
to a change of these reference wavelengths and frequencies over cosmological
distances and times. We will adopt this possible interpretation in the
following, altering the classical meaning of $\nu$, $\nu_{0}$ and $\lambda$,
$\lambda_{0}$ in Eq. (\ref{eqn3.16}).

In our alternative redshift interpretation we assume that the observed
quantities $\lambda_{0}$ and $\nu_{0}$, are telling us about the radiation
emitted by the source galaxy at the place and time of emission, while the
reference quantities $\lambda$ and $\nu$ are, by the same argument,
characteristics of the same atomic radiation as measured here on Earth at
present time. We therefore write:%

\begin{align}
\lambda_{0}  &  =\lambda(\mathbf{r,t})\ ;\ \nu_{0}=\nu(\mathbf{r,t}%
)\label{eqn3.18}\\
\lambda &  =\lambda(\mathbf{0,t}_{0})\ ;\ \nu=\nu(\mathbf{0,t}_{0})\nonumber
\end{align}
where again $\mathbf{r=0}$ represents the Earth's observer position,
$\mathbf{t}_{0}$ is the present time, while $\mathbf{r}$ is the position of
the source galaxy and $\mathbf{t}$ is the time of emission, in the past. Since
the units of length and time, defined in Eq. (\ref{eqn3.17}), are respectively
proportional to the radiation wavelengths and inversely proportional to the
radiation frequencies, they also become functions of the space-time coordinates:%

\begin{align}
1\ meter  &  \equiv\delta l(\mathbf{r,t})=n_{m}\ \lambda_{m}(\mathbf{r,t}%
)\label{eqn3.19}\\
1\ second  &  \equiv\delta t(\mathbf{r,t})=\frac{n_{s}}{\nu_{s}(\mathbf{r,t}%
)},\nonumber
\end{align}
where $\delta l$ and $\delta t$ will indicate the unit-length and the
unit-time-interval\ in the following. Due to this new interpretation, we
correct Eq. (\ref{eqn3.16}), combining it also with the previous equations:%

\begin{equation}
1+z=\frac{\mathbf{R(t}_{0}\mathbf{)}}{\mathbf{R(t)}}=\frac{\lambda
(\mathbf{r,t})}{\lambda(\mathbf{0,t}_{0})}=\frac{\delta l(\mathbf{r,t}%
)}{\delta l(\mathbf{0,t}_{0})}=\frac{\nu(\mathbf{0,t}_{0})}{\nu(\mathbf{r,t}%
)}=\frac{\delta t(\mathbf{r,t})}{\delta t(\mathbf{0,t}_{0})}. \label{eqn3.20}%
\end{equation}

In view of the modern definition of the unit of length, based on a fixed value
of $c$, we will consider the value of the speed of light in vacuum to be just
a connecting factor between the units of length and time and we see no reason,
at least for now, to assume that this factor might also change at different
space-time locations. In our opinion, a variation of the speed of light
(proposed by some alternative cosmologies \cite{Albrecht:1998ir},
\cite{Barrow:1999is}, \cite{Magueijo:2000zt}) would imply a substantial
difference in the universal evolution of the units of length and time which
seems an unnecessary complication, not supported by experimental observations.
Therefore, we will consider $c=2.99792458\times10^{10}\ \frac{cm}{s}$ as a
constant value in the following, but we will continue to explicitly include
$c$ in every equation.

\section{\label{sect:evaluation_cosmic}Evaluation of the Cosmic Scale Factor}

The alternative interpretation of the cosmological redshift, presented in the
previous section, is actually an adaptation of the well-known
\textit{gravitational redshift} (or \textit{gravitational time-dilation}) to
the cosmological scale and was even considered in the 1920's as a possible
origin of the observed redshift (see the historical discussion in Weinberg
\cite{Weinberg}, page 417), but would have required very strong local
gravitational fields, so this explanation was quickly abandoned in favor of a
\textquotedblleft cosmological\textquotedblright\ Doppler effect.
Nevertheless, it is interesting to notice that this possibility was taken into
account at the beginning of modern cosmology as well as many other explanations.

The gravitational redshift is a fundamental consequence of the equivalence
principle which states that the rate of a clock at rest is affected by the
presence of a gravitational field as follows:%

\begin{equation}
\frac{\delta t}{\Delta t}=\frac{1}{\sqrt{-g_{00}(x)}}, \label{eqn4.1}%
\end{equation}
where $\delta t$ and $\Delta t$ are respectively the clock periods in the
presence or in the absence of gravitation, and $g_{00}(x)$ is the value of the
time component of the metric at the point where the clock is located (see
\cite{Weinberg}, section 3.5 for a general discussion, or \cite{Lang:1999ry}
for a review of experimental results). Since the \textquotedblleft
true\textquotedblright\ period $\Delta t$ of a clock is unknown, we can only
observe this effect by comparing the rate of the clock at two different
locations $x_{1}$, $x_{2}$ in the gravitational field:%

\begin{equation}
\frac{\delta t_{1}}{\delta t_{2}}=\sqrt{\frac{g_{00}(x_{2})}{g_{00}(x_{1})}%
}=\frac{\nu_{2}}{\nu_{1}}=\frac{\lambda_{1}}{\lambda_{2}}\equiv1+z
\label{eqn4.2}%
\end{equation}
and this quantity is related to the ratio of the frequencies or wavelengths of
the same atomic transition observed at the two locations, which can also be
described by a \textquotedblleft redshift\textquotedblright\ parameter $z$.
The connection between Eqs. (\ref{eqn3.20}) and (\ref{eqn4.2}) is immediate,
identifying the two locations $x_{1}$, $x_{2}$ with $(\mathbf{r,t})$ and
$(\mathbf{0,t}_{0})$ respectively.

The \textit{gravitational redshift} or \textit{time dilation} has been tested
repeatedly for the classic Schwarzschild solution of the metric, i.e.,
$g_{00}(x)=-B(r)=-(1-\frac{2MG}{c^{2}r})$. Using this expression inside Eq.
(\ref{eqn4.2}) we obtain a redshift if the point of emission\ $x_{1}$ is
closer to the massive source of the field, compared to the point of
observation\ $x_{2}$, such as in the case of light emitted by the Sun or by
white dwarfs and observed here on Earth \cite{Lang:1999ry}. A blueshift can be
observed instead by using the Earth's gravity and by placing point $x_{1}$ at
a higher level\ than point $x_{2}$, as in the classic experiment by Pound and
Rebka (see description in \cite{Weinberg}). These gravitational redshifts are
very small (the one due to the Sun corresponds to $z\sim10^{-6}$ and those
related to white dwarfs are about two orders of magnitude bigger) and cannot
produce any cosmological redshift, since they are just a local effect,
predicted on the basis of the classic Schwarzschild solution for a static and
spherically\ symmetric massive source, such as a planet or a star.

However, in view of the preceding discussion of the cosmological redshift and
of the new MK solutions shown in Eq. (\ref{eqn2.8}) or Eq. (\ref{eqn3.1}),
involving a cosmological generalization of the classic Schwarzschild solution
through the cosmological parameters $\gamma$ and $\kappa$, we can now propose
a direct determination of the scale factor $\mathbf{R(t)}$ based on this
\textquotedblleft extended\textquotedblright\ interpretation of the
\textit{gravitational-cosmological redshift}. In other words, we will show
that, assuming the validity of Conformal Gravity and of the interpolation
between the Static Standard Coordinates and the Robertson-Walker
metric\ explained in Sect. \ref{sect:from_SSC_to_RW}, our alternative redshift
interpretation restricts the possible conformal transformations of the metric
to just one possible case, i.e., just one possible function $\Omega(x)$ in Eq.
(\ref{eqn2.2}), therefore also practically breaking this conformal symmetry
without resorting to field-theory symmetry breaking procedures.

The function $\mathbf{R(t)}$ will be uniquely determined from these purely
\textquotedblleft kinematical\textquotedblright\ considerations and we will
not need to obtain it from the solution of the \textquotedblleft
dynamic\textquotedblright\ field equation (\ref{eqn2.4}) of conformal gravity,
as it is done in standard General Relativity using the Friedmann equations. We
will then compare our solutions for $\mathbf{R(t)}$ with the corresponding
solutions obtained by current conformal cosmologies in the literature, since
the metric in Eqs. (\ref{eqn2.7})-(\ref{eqn2.8}) is based on the same equation
of motion of conformal gravity, i.e., Eq. (\ref{eqn2.4}) with $T_{\mu\nu}=0$.

\subsection{\label{sect:function_of_distance}The cosmic scale factor as a
function of the radial coordinates}

It is immediate to obtain the cosmic scale factor as a function of the radial
coordinates. We start by combining Eqs. (\ref{eqn3.1}) and (\ref{eqn3.13}), in
order to rewrite $B(r)$ as:%
\begin{equation}
B(r)=1+\gamma r+\left(  \frac{\gamma^{2}}{4}+k\right)  r^{2}=-g_{00}(r)
\label{eqn4.3}%
\end{equation}
in Static Standard Coordinates. Now we use Eq. (\ref{eqn4.2}) to compute the
\textit{gravitational-cosmological} time dilation for two points corresponding
to the source galaxy space-time position and the Earth's observer placed at
the origin at present time:%

\begin{equation}
1+z=\frac{R(0)}{R(r)}=\frac{\lambda(r,t)}{\lambda(0,t_{0})}=\frac{\nu
(0,t_{0})}{\nu(r,t)}=\sqrt{\frac{-g_{00}(0)}{-g_{00}(r)}}=\frac{1}%
{\sqrt{1+\gamma r+\left(  \frac{\gamma^{2}}{4}+k\right)  r^{2}}},
\label{eqn4.4}%
\end{equation}
which gives the redshift factor $\left(  1+z\right)  $ as a very simple
function of the coordinate $r$ in SSC. We also express the factor $\left(
1+z\right)  $\ as a ratio of cosmic scale factors, computed at the two points
of interest, although usually the scale factor is only introduced in the RW
metric. We will show in the following that this function $R$ can be expressed
in any of the space-time coordinates of interest, therefore we can also
introduce it in the SSC.

Our objective is now to transform this expression into RW coordinates, by
using the transformations outlined in Sect. \ref{sect:from_SSC_to_RW}. Before
doing this, we note that Eq. (\ref{eqn4.4}) should give the observed
cosmological redshift (i.e., $1+z>1$) at least for some distance interval
$r>r_{rs}$, where $r_{rs}$\ is the coordinate beyond which we start observing
a cosmological redshift. In addition, we assume that the parameter $\gamma$ is
small and positive at the present time, probably close to Mannheim's value of
$\gamma_{0}\simeq3.06\times10^{-30}%
\operatorname{cm}%
^{-1}$, while the parameter $k$ is not yet restricted ($k\gtreqqless0$).

For $\gamma>0$, a quick inspection of Eq. (\ref{eqn4.4}) shows that a solution
allowing redshift is possible only in one case: for a negative $k$ and more
precisely for $k<-\frac{\gamma^{2}}{4}$. In this case the function in Eq.
(\ref{eqn4.4}) is well defined for positive values of $r$ in the interval
$0\leq r<1/(\sqrt{\left\vert k\right\vert }-\frac{\gamma}{2})$. Moreover, we obtain:%

\begin{equation}
r_{rs}=\gamma/(\left\vert k\right\vert -\frac{\gamma^{2}}{4}), \label{eqn4.5}%
\end{equation}
giving a blueshift ($z<0$) for distances in the interval $0<r<r_{rs}$, and a
proper redshift ($z>0$) for larger distances $r>r_{rs}$, which might
correspond to the observed cosmological redshift.

Since the cosmic scale factor and all the other cosmological quantities of
interest are usually expressed in Robertson-Walker coordinates, we have to
convert the expression in Eq. (\ref{eqn4.4}) into these coordinate. This can
be accomplished by using the transformations of Sect.
\ref{sect:from_SSC_to_RW}. From Eq. (\ref{eqn3.3}) and its inverse
transformation, it follows that%

\begin{equation}
\left[  1+\gamma r+\left(  \frac{\gamma^{2}}{4}+k\right)  r^{2}\right]
=\frac{[1+\frac{k}{4}\rho^{2}]^{2}}{[1-\frac{\gamma}{2}\rho-\frac{k}{4}%
\rho^{2}]^{2}}, \label{eqn4.6}%
\end{equation}
so that we can write%

\begin{equation}
1+z=\frac{1}{\sqrt{1+\gamma r+\left(  \frac{\gamma^{2}}{4}+k\right)  r^{2}}%
}=\frac{[1-\frac{\gamma}{2}\rho-\frac{k}{4}\rho^{2}]}{[1+\frac{k}{4}\rho^{2}%
]}, \label{eqn4.7}%
\end{equation}
which is well defined for $0\leq\rho<2/\sqrt{\left\vert k\right\vert }$.

The conformal transformation of Eq. (\ref{eqn3.8})\ will not alter the $\rho$
coordinate, so we just need to apply the final two transformations of Eqs.
(\ref{eqn3.11}) and (\ref{eqn3.14}) to obtain, after some algebraic work:%

\begin{equation}
1+z=\frac{\mathbf{R(0)}}{\mathbf{R(r)}}=\sqrt{1-k\rho^{\prime2}}-\frac{\gamma
}{2}\rho^{\prime}=\sqrt{1-\mathbf{k\ r}^{2}}-\frac{\gamma}{2\sqrt{\left\vert
k\right\vert }}\mathbf{r,} \label{eqn4.8}%
\end{equation}
where we use again $\mathbf{r}$ (in bold) to denote the radial coordinate in
RW metric and $\mathbf{k}=\frac{k}{\left\vert k\right\vert }=0,\pm1$,
following Eq. (\ref{eqn3.14}). We observe that the last term in the previous
equation diverges for $k=0$, but according to the note following Eq.
(\ref{eqn3.14}) in this particular case the previous equation should simply
become $1+z=\frac{\mathbf{R(0)}}{\mathbf{R(r)}}=1-\frac{\gamma}{2}\rho
^{\prime}=1-\frac{\gamma}{2}\mathbf{r}$. Another way to obtain Eq.
(\ref{eqn4.8}) from Eq. (\ref{eqn4.4}) is to use the direct connection between
coordinates $r$ and $\mathbf{r}$, which can be easily derived from the
transformations of Sect. \ref{sect:from_SSC_to_RW} and is the following:%

\begin{equation}
\sqrt{\left\vert k\right\vert }r=\left(  \sqrt{\frac{1}{\mathbf{r}^{2}%
}-\mathbf{k\ }}-\frac{\gamma}{2\sqrt{\left\vert k\right\vert }}\right)  ^{-1}.
\label{eqn4.9}%
\end{equation}

A more elegant way to write the previous fundamental equations is to introduce
a dimensionless parameter:%

\begin{equation}
\delta\equiv\frac{\gamma}{2\sqrt{\left\vert k\right\vert }}, \label{eqn4.10}%
\end{equation}
or $\delta\equiv\frac{\gamma}{2}$ for the particular case $k=0$, and rewrite
Eq. (\ref{eqn4.8}) as%

\begin{equation}
1+z=\frac{\mathbf{R(0)}}{\mathbf{R(r)}}=\sqrt{1-\mathbf{k\ r}^{2}}%
-\delta\mathbf{r\ ;\ \ k}=0,\pm1. \label{eqn4.11}%
\end{equation}
We have written the ratio of cosmic scale factors as a function of the radial
coordinates of the points of emission and absorption of radiation, since the
function on the right-hand side of the previous equation depends only on
$\mathbf{r}$, although we are implicitly referring also to the times at which
the radiation was emitted and absorbed. A more precise notation would be to
write always these cosmological scale factors as $\mathbf{R(r,t)}$ and
$\mathbf{R(0,t}_{0}\mathbf{)}$ in all our formulas, but we will continue to
use our simplified notation also in the following.

In the last equations the parameter $\gamma$ is positive and determined, at
least for now, by Mannheim's fits of galactic rotational curves, while the
other parameter $k=-\frac{\gamma^{2}}{4}-\kappa$ is still undetermined. Since
a cosmological redshift is generally observed,\footnote{Except for some nearby
galaxies typically located in our Local Group, whose blueshift is presumably
due to their peculiar motion, or for the \textquotedblleft Pioneer anomalous
blueshift\textquotedblright\ which will be considered in another paper
\cite{Varieschi:2008va}.} i.e., $\mathbf{R(0)/R(r)}>0$ in general, we have
already remarked that this suggests a negative value of $k<-\frac{\gamma^{2}%
}{4}<0$, which implies $0<\delta^{2}\equiv\frac{\gamma^{2}}{4\left\vert
k\right\vert }<1$, thus restricting in general $\delta$ to the interval
$-1<\delta<1$ (but with a currently positive value).

For these values of the parameters the quantity $r_{rs}=\gamma/(\left\vert
k\right\vert -\frac{\gamma^{2}}{4})$ introduced in Eq. (\ref{eqn4.5}) can be
rewritten in RW coordinates as%

\begin{equation}
\mathbf{r}_{rs}=1/\left(  \sqrt{\left\vert k\right\vert }/\gamma-\gamma
/4\sqrt{\left\vert k\right\vert }\right)  =\frac{2\delta}{1-\delta^{2}}.
\label{eqn4.12}%
\end{equation}

Again, we have a standard redshift for distances $\mathbf{r}>\mathbf{r}_{rs}$,
but an unexpected blueshift at closer distances $0<\mathbf{r}<\mathbf{r}_{rs}%
$. This possibility is particularly interesting in view of a recently
discovered phenomenon, the so-called \textit{Pioneer anomaly}
(\cite{Anderson:1998jd}, \cite{Anderson:2001sg}, \cite{Turyshev:2005zk},
\cite{Turyshev:2005ej}, \cite{Toth:2006qb}) consisting in an anomalous
blueshift observed in the navigation of the Pioneer spacecraft, just outside
the Solar System.

\subsection{\label{sect:time_dependent_form}Time-dependent form of the cosmic
scale factor}

The cosmic scale factor $\mathbf{R}$ is usually considered a function of some
cosmic time coordinate $\mathbf{t}$, rather than a function of the radial
coordinates, as introduced in the previous section. This is a consequence of
the Cosmological Principle (i.e., the Universe being assumed spatially
homogeneous and isotropic) and of the application of this principle to the
hypersurfaces with constant cosmic standard time, which are maximally
symmetric subspaces of the whole of space-time (see Chapters 13-14 in Ref.
\cite{Weinberg} for details). Following this standard hypothesis, the
resulting metric takes the RW form of Eq. (\ref{eqn3.15}) and the redshift is
described by the ratio of scale factors at two different cosmic times, as in
Eq. (\ref{eqn3.16}).

On the contrary, our new interpretation assumes that the redshift is due to
the stretching of the space-time fabric as described by Eqs. (\ref{eqn4.4})
and (\ref{eqn4.8}), which are essentially static solutions, derived from the
Conformal Gravity theory. In order to retain the validity of the Cosmological
Principle and, in particular, still assume the homogeneity of the Universe at
a given cosmic time, we have to transform the space dependence of our new
cosmic scale factors into a more traditional time dependence.

This can be accomplished by noting that the redshifted radiation described by
Eqs. (\ref{eqn4.4}) or (\ref{eqn4.8}) is reaching us from past times and that
light coming from a radial distance $\mathbf{r}$ is all emitted at the same
time $\mathbf{t}$ in the past. Therefore, the scale factor $\mathbf{R(r)}$ can
be associated with a corresponding factor $\mathbf{R(t)}$, at a given past
cosmic time $\mathbf{t}$. This association is performed by computing the time
it takes for a light signal emitted at radial distance $\mathbf{r}$ to reach
the observer at the origin. It is then straightforward to turn Eqs.
(\ref{eqn4.4}) or (\ref{eqn4.8}) into their time dependent equivalent, since
we are following a light signal traveling in vacuum from a distant galaxy
toward us, for which $ds^{2}=0$ or $d\widehat{s}^{2}=0$.

It is convenient to study the propagation of this light signal by using first
our original Static Standard Coordinates $(r,t,\theta,\phi)$. Combining Eqs.
(\ref{eqn3.2}) and (\ref{eqn4.3}), for a light ray traveling along the $(-r)$
direction with $\theta$ and $\phi$ fixed, we have:%

\begin{equation}
ds^{2}=-[1+\gamma r+(\frac{\gamma^{2}}{4}+k)\ r^{2}]\ c^{2}dt^{2}+\frac
{dr^{2}}{[1+\gamma r+(\frac{\gamma^{2}}{4}+k)\ r^{2}]}=0, \label{eqn4.13}%
\end{equation}
or equivalently%

\begin{equation}
\ c\ dt=-\frac{dr}{[1+\gamma r+(\frac{\gamma^{2}}{4}+k)\ r^{2}]},
\label{eqn4.14}%
\end{equation}
where the negative sign comes from the $(-r)$ direction of light propagation.

Integrating between times $t$ and $t_{0}$, corresponding to radial positions
$r$ and $r=0$, we obtain different results, depending on the sign of the
parameter $k$:%

\begin{align}
x  &  \equiv c\ (t_{0}-t)=\int_{0}^{r}\frac{dr^{\prime}}{[1+\gamma r^{\prime
}+(\frac{\gamma^{2}}{4}+k)\ r^{\prime2}]}=\label{eqn4.15}\\
&  =\left[  \frac{1}{\sqrt{k}}\tan^{-1}\left(  \frac{4kr^{\prime}%
+2\gamma+\gamma^{2}r^{\prime}}{4\sqrt{k}}\right)  \right]  _{0}^{r}=\frac
{i}{2\sqrt{k}}\ln\left[  \frac{1+(\frac{\gamma}{2}-i\sqrt{k})\ r}%
{1+(\frac{\gamma}{2}+i\sqrt{k})\ r}\right]  \ ;\ k>0\nonumber\\
&  =\left[  \frac{-4}{\gamma(2+\gamma r^{\prime})}\right]  _{0}^{r}=\frac
{r}{(1+\frac{\gamma}{2}\ r)}\ ;\ k=0\nonumber\\
&  =\left[  \frac{1}{\sqrt{\left\vert k\right\vert }}\tanh^{-1}\left(
\frac{4\left\vert k\right\vert r^{\prime}-2\gamma-\gamma^{2}r^{\prime}}%
{4\sqrt{\left\vert k\right\vert }}\right)  \right]  _{0}^{r}=\frac{1}%
{2\sqrt{\left\vert k\right\vert }}\ln\left[  \frac{1+(\frac{\gamma}{2}%
+\sqrt{\left\vert k\right\vert })\ r}{1+(\frac{\gamma}{2}-\sqrt{\left\vert
k\right\vert })\ r}\right]  \ ;\ k<0,\nonumber
\end{align}
where we have introduced the useful quantity $x\equiv c\ (t_{0}-t)$. It is
possible to invert Eq. (\ref{eqn4.15}) in each case and obtain the distance
$r$ as a function of $x\equiv c\ (t_{0}-t)$:%

\begin{align}
r  &  =\frac{1}{\sqrt{k}}\ \frac{1}{\left[  \cot\left(  \sqrt{k}x\right)
-\frac{\gamma}{2\sqrt{k}}\right]  }\ ;\ k>0\label{eqn4.16}\\
r  &  =\ \frac{x}{\left[  1-\frac{\gamma}{2}x\right]  }\ ;\ k=0\nonumber\\
r  &  =\frac{1}{\sqrt{\left\vert k\right\vert }}\ \frac{1}{\left[
\coth\left(  \sqrt{\left\vert k\right\vert }x\right)  -\frac{\gamma}%
{2\sqrt{\left\vert k\right\vert }}\right]  }\ ;\ k<0.\nonumber
\end{align}

Finally, it takes a little more work to combine Eq. (\ref{eqn4.16}) together
with Eq. (\ref{eqn4.4}), to obtain the explicit form of the cosmic scale factor:%

\begin{align}
1+z  &  =\frac{R(t_{0})}{R(t)}=\left[  \cos\left(  \sqrt{k}x\right)
-\frac{\gamma}{2\sqrt{k}}\sin\left(  \sqrt{k}x\right)  \right]
\ ;\ k>0\label{eqn4.17}\\
1+z  &  =\frac{R(t_{0})}{R(t)}=\left[  1-\frac{\gamma}{2}x\right]
\ ;\ k=0\nonumber\\
1+z  &  =\frac{R(t_{0})}{R(t)}=\left[  \cosh\left(  \sqrt{\left\vert
k\right\vert }x\right)  -\frac{\gamma}{2\sqrt{\left\vert k\right\vert }}%
\sinh\left(  \sqrt{\left\vert k\right\vert }x\right)  \right]
\ ;\ k<0,\nonumber
\end{align}
a remarkably compact expression in each case. To obtain the scale factor as a
function of the cosmic time coordinate $\mathbf{t}$ we could repeat the same
procedure, studying the propagation of light in RW metric, but this involves
rather cumbersome integrals. It is easier to find a direct relation between
the time coordinates $t$ and $\mathbf{t}$.

We start combining together Eqs. (\ref{eqn3.3}), (\ref{eqn3.11}) and
(\ref{eqn3.14}), obtaining%

\begin{equation}
d\mathbf{t}=R(t)\ dt \label{eqn4.18}%
\end{equation}
or, integrating between times $t$, $\mathbf{t}$ in the past and present times
$t_{0}$, $\mathbf{t}_{0}$,%

\begin{equation}
\mathbf{t}_{0}-\mathbf{t}=%
{\displaystyle\int_{t}^{t_{0}}}
R(t)\ dt \label{eqn4.19}%
\end{equation}
where the cosmic scale factor $R(t)$ is expressed through Eq. (\ref{eqn4.17}):%

\begin{align}
R(t)  &  =R(t_{0})/\left[  \cos\left(  \sqrt{k}x\right)  -\frac{\gamma}%
{2\sqrt{k}}\sin\left(  \sqrt{k}x\right)  \right]  \ ;\ k>0\label{eqn4.20}\\
R(t)  &  =R(t_{0})/\left[  1-\frac{\gamma}{2}x\right]  \ ;\ k=0\nonumber\\
R(t)  &  =R(t_{0})/\left[  \cosh\left(  \sqrt{\left\vert k\right\vert
}x\right)  -\frac{\gamma}{2\sqrt{\left\vert k\right\vert }}\sinh\left(
\sqrt{\left\vert k\right\vert }x\right)  \right]  \ ;\ k<0.\nonumber
\end{align}
Integrating Eq. (\ref{eqn4.19}) with the expressions from Eq. (\ref{eqn4.20})
we obtain (using the more compact parameter $\delta\equiv\gamma/2\sqrt
{\left\vert k\right\vert }$):%

\begin{align}
\mathbf{t}_{0}-\mathbf{t}  &  =\frac{2\mathbf{R(t}_{0}\mathbf{)}}%
{c\sqrt{1+\delta^{2}}}\operatorname{arccoth}\left[  \frac{\cot\left(
\frac{\sqrt{k}}{2}x\right)  -\delta}{\sqrt{1+\delta^{2}}}\right]
\ ;\ k>0\label{eqn4.21}\\
\mathbf{t}_{0}-\mathbf{t}  &  =-\frac{2\mathbf{R(t}_{0}\mathbf{)}}{c\gamma}%
\ln\left[  1-\frac{\gamma}{2}x\right]  \ ;\ k=0\nonumber\\
\mathbf{t}_{0}-\mathbf{t}  &  =\frac{2\mathbf{R(t}_{0}\mathbf{)}}%
{c\sqrt{1-\delta^{2}}}\operatorname{arccot}\left[  \frac{\coth\left(
\frac{\sqrt{\left\vert k\right\vert }}{2}x\right)  -\delta}{\sqrt{1-\delta
^{2}}}\right]  \ ;\ k<0.\nonumber
\end{align}
Inverting these expressions, we obtain the connections between the time coordinates:%

\begin{align}
x  &  \equiv c\ (t_{0}-t)=\frac{2}{\sqrt{k}}\operatorname{arccot}\left\{
\sqrt{1+\delta^{2}}\coth\left[  \frac{\sqrt{1+\delta^{2}}}{2}\frac{c\left(
\mathbf{t}_{0}-\mathbf{t}\right)  }{\mathbf{R(t}_{0}\mathbf{)}}\right]
+\delta\right\}  \ ;\ k>0\label{eqn4.22}\\
x  &  \equiv c\ (t_{0}-t)=\frac{2}{\gamma}\left[  1-e^{-\frac{\gamma}{2}%
\frac{c\left(  \mathbf{t}_{0}-\mathbf{t}\right)  }{\mathbf{R(t}_{0}\mathbf{)}%
}}\right]  \ ;\ k=0\nonumber\\
x  &  \equiv c\ (t_{0}-t)=\frac{2}{\sqrt{\left\vert k\right\vert }%
}\operatorname{arccoth}\left\{  \sqrt{1-\delta^{2}}\cot\left[  \frac
{\sqrt{1-\delta^{2}}}{2}\frac{c\left(  \mathbf{t}_{0}-\mathbf{t}\right)
}{\mathbf{R(t}_{0}\mathbf{)}}\right]  +\delta\right\}  \ ;k<0.\nonumber
\end{align}
Inserting this last equation into Eq. (\ref{eqn4.20}), we finally compute the
cosmic scale factor in RW coordinates:%

\begin{align}
\mathbf{R(t)}  &  =\mathbf{R(t}_{0}\mathbf{)}\ \left\{  \cosh\left[
\sqrt{1+\delta^{2}}\frac{c(\mathbf{t}_{0}-\mathbf{t)}}{\mathbf{R(t}%
_{0}\mathbf{)}}\right]  +\frac{\delta}{\sqrt{1+\delta^{2}}}\sinh\left[
\sqrt{1+\delta^{2}}\frac{c(\mathbf{t}_{0}-\mathbf{t)}}{\mathbf{R(t}%
_{0}\mathbf{)}}\right]  \right\}  ;\ \mathbf{k}=1\label{eqn4.23}\\
\mathbf{R(t)}  &  =\mathbf{R(t}_{0}\mathbf{)}\ e^{\frac{\gamma}{2}%
\frac{c\left(  \mathbf{t}_{0}-\mathbf{t}\right)  }{\mathbf{R(t}_{0}\mathbf{)}%
}}\ ;\ \mathbf{k}=0\nonumber\\
\mathbf{R(t)}  &  =\mathbf{R(t}_{0}\mathbf{)}\ \left\{  \cos\left[
\sqrt{1-\delta^{2}}\frac{c\left(  \mathbf{t}_{0}-\mathbf{t}\right)
}{\mathbf{R(t}_{0}\mathbf{)}}\right]  +\frac{\delta}{\sqrt{1-\delta^{2}}}%
\sin\left[  \sqrt{1-\delta^{2}}\frac{c\left(  \mathbf{t}_{0}-\mathbf{t}%
\right)  }{\mathbf{R(t}_{0}\mathbf{)}}\right]  \right\}  ;\ \mathbf{k}%
=-1.\nonumber
\end{align}

We will study all these functions in detail in the following sections. We
remark here that we now have the cosmological scale factor expressed in four
different types of coordinates. The space coordinates $r$ and $\mathbf{r}$,
entering Eqs. (\ref{eqn4.4}) and (\ref{eqn4.11}) respectively, are appropriate
to measure distances in their respective coordinate systems: $r$ (in
centimeters or meters) refers to the Static Standard Coordinates and to the
local \textquotedblleft meter stick\textquotedblright\ being used, it is
therefore suitable for local measurements. To measure distances on
astronomical scales we will need to introduce revised expressions for the
classic luminosity distance and for the other distances used in cosmology,
which are all based on the Robertson-Walker $\mathbf{r}$ (dimensionless)
coordinate. Therefore, when measuring distances to galaxies, supernovae, etc.
(in $Mpc$, light years, or other suitable units) we will use the $\mathbf{r}%
$\ coordinate and our fundamental cosmological scale factor will be given by
Eq. (\ref{eqn4.11}).

A different situation exists for the two time coordinates $t$ and $\mathbf{t}%
$, or for the equivalent look-back times $t_{0}-t$\ and $\mathbf{t}%
_{0}-\mathbf{t}$. Although we measure both of them with the same units
(seconds, years, etc.), the time interval between two events is expressed
differently by the two time coordinates, as shown in Eqs. (\ref{eqn4.21}) and
(\ref{eqn4.22}) or by the original connection in Eq. (\ref{eqn4.19}). In
standard cosmology, when making time determinations such as estimating the age
of the Universe or using radioactive decay for the age determination of
astronomical objects, it is assumed that our local time $t$ can be
synchronized with the cosmological time $\mathbf{t}$, therefore no distinction
is made between the two.

However, in our model these two quantities are different and their connection
is given above. When using the comoving RW coordinates, one should refer to
the cosmic time $\mathbf{t}$ (in bold) and use Eq. (\ref{eqn4.23}). On the
contrary, when making contact with observations, such as age estimates or when
taking time derivatives of the cosmic scale factor to obtain the Hubble and
deceleration parameters, the local time coordinate $t$ should be used,
together with Eq. (\ref{eqn4.17}), since all these observations refer to our
terrestrial clocks.

We now compare our solutions in Eq. (\ref{eqn4.23}) with the corresponding
solutions obtained by current conformal cosmologies in the literature, as we
already remarked that the metric in Eqs. (\ref{eqn2.7})-(\ref{eqn2.8}) is
based on the same equation of motion of conformal gravity, i.e., Eq.
(\ref{eqn2.4}) with $T_{\mu\nu}=0$.

Our solutions in Eq. (\ref{eqn4.23}) are similar to those proposed by Mannheim
in Eq. (9) of Ref. \cite{Mannheim:1999nc} and also discussed in Ref.
\cite{Mannheim:1998ew}. In particular, Mannheim analyzes the \textquotedblleft
de Sitter geometry in a purely kinematic way that requires no commitment to
any particular dynamical equation of motion, neither that of conformal gravity
nor that of the standard model\textquotedblright\ (quoted from
\cite{Mannheim:1999nc}) and it is therefore related to a cosmology with just a
cosmological constant source. In Robertson-Walker coordinates a simple
kinematic relation is then obtained (see Eq. (8) of Ref.
\cite{Mannheim:1999nc}):%

\begin{equation}
\mathbf{\overset{\cdot}{\mathbf{R}}{}^{2}}(\mathbf{\mathbf{t}})+\mathbf{k}%
c^{2}=\alpha c^{2}\mathbf{R}^{2}(\mathbf{t}) \label{eqn4.23.1}%
\end{equation}
and five possible cosmological solutions (see Eq. (9) of Ref.
\cite{Mannheim:1999nc}) are found by Mannheim, for the different cases related
to the signs of the parameters $\alpha$ and $\mathbf{k}$:%

\begin{align}
\mathbf{R(t,}\alpha &  <0\mathbf{,k}<0\mathbf{)}=\left(  \frac{\mathbf{k}%
}{\alpha}\right)  ^{1/2}\sin\left[  \left(  -\alpha\right)  ^{1/2}%
c\mathbf{t}\right] \label{eqn4.23.2}\\
\mathbf{R(t,}\alpha &  =0\mathbf{,k}<0\mathbf{)}=\left(  -\mathbf{k}\right)
^{1/2}c\mathbf{t}\nonumber\\
\mathbf{R(t,}\alpha &  >0\mathbf{,k}<0\mathbf{)}=\left(  -\frac{\mathbf{k}%
}{\alpha}\right)  ^{1/2}\sinh\left(  \alpha^{1/2}c\mathbf{t}\right)
\nonumber\\
\mathbf{R(t,}\alpha &  >0\mathbf{,k}=0\mathbf{)=R(t}=0\mathbf{)}\exp\left(
\alpha^{1/2}c\mathbf{t}\right) \nonumber\\
\mathbf{R(t,}\alpha &  >0\mathbf{,k}>0\mathbf{)=}\left(  \frac{\mathbf{k}%
}{\alpha}\right)  ^{1/2}\cosh\left(  \alpha^{1/2}c\mathbf{t}\right)
.\nonumber
\end{align}

We can rewrite our three solutions in Eq. (\ref{eqn4.23}) so that they are
equivalent to three of Mannheim's expressions in the previous equation, as
long as we identify the time variables as follows: $\left(  \mathbf{t}%
_{0}-\mathbf{t}\right)  \mathbf{\rightarrow t}$ and $\mathbf{t}_{0}%
\rightarrow\left(  \mathbf{t}=0\right)  $. Our Eq. (\ref{eqn4.23}) then becomes:%

\begin{align}
\mathbf{R(t,\mathbf{k}}  &  =1\mathbf{)}=\frac{\mathbf{R(\mathbf{t}%
}=0\mathbf{)}}{\sqrt{1+\delta^{2}}}\ \cosh\left[  \epsilon+\frac
{\sqrt{1+\delta^{2}}}{\mathbf{R(\mathbf{t}}=0\mathbf{)}}c\mathbf{t}\right]
;\text{ with }\epsilon=\sinh^{-1}\delta\label{eqn4.23.3}\\
\mathbf{R(t,\mathbf{k}}  &  =0\mathbf{)}=\mathbf{R(t}=0\mathbf{)}\ \exp\left[
\frac{\gamma}{2\mathbf{R(t}=0\mathbf{)}}c\mathbf{t}\right] \nonumber\\
\mathbf{R(t,\mathbf{k}}  &  =-1\mathbf{)}=\frac{\mathbf{R(\mathbf{t}%
}=0\mathbf{)}}{\sqrt{1-\delta^{2}}}\ \sin\left[  \epsilon+\frac{\sqrt
{1-\delta^{2}}}{\mathbf{R(\mathbf{t}}=0\mathbf{)}}c\mathbf{t}\right]  ;\text{
with }\epsilon=\cos^{-1}\delta.\nonumber
\end{align}

These three solutions correspond respectively to the fifth, fourth and first
expression in Eq. (\ref{eqn4.23.2}), with an additional quantity denoted by
$\epsilon$, in the first and third solution, which depends on our parameter
$\delta$. It can be easily checked that all three expressions in Eq.
(\ref{eqn4.23.3}) verify the original kinematic relation in Eq.
(\ref{eqn4.23.1}), with $\alpha=\frac{1+\delta^{2}}{\mathbf{R}^{2}%
\mathbf{(\mathbf{t}}=0\mathbf{)}};\frac{\gamma^{2}}{4\mathbf{R}^{2}%
\mathbf{(\mathbf{t}}=0\mathbf{)}};-\frac{1-\delta^{2}}{\mathbf{R}%
^{2}\mathbf{(\mathbf{t}}=0\mathbf{)}}$ respectively in the three cases.
Therefore, our fundamental expressions in Eq. (\ref{eqn4.23}) are valid
solutions of the problem analyzed by Mannheim and described above.

Following this discussion, it can be noted \cite{Reviewer:1} that the
solutions we obtained with our kinematical approach to conformal cosmology are
equivalent to those of the standard conformal cosmology
(\cite{Mannheim:1998ew}, \cite{Mannheim:1999nc}) based solely on a
cosmological constant. In this view, our approach could be valid in two
separate epochs: a very early Universe which undergoes a cosmological constant
dominated inflationary phase and a very late Universe in which the
cosmological constant dominates the energy density.

The identification of the cosmological redshift with the gravitational
redshift could then represent not the current state of the Universe, but
rather the one into which the Universe might ultimately evolve, or the one
that it may initially have evolved from \cite{Reviewer:1}. Therefore, it is
noteworthy that our kinematic cosmology is actually recovered from conformal
gravity with a cosmological constant at late (or early) times and this result
constitutes the central point of this paper.

Finally, in addition to the connecting relations between $r$, $\mathbf{r}$,
$t$ and $\mathbf{t}$ given by Eqs. (\ref{eqn4.9}), (\ref{eqn4.15}),
(\ref{eqn4.16}), (\ref{eqn4.21}) and (\ref{eqn4.22}), a very simple connection
exists between the RW $\mathbf{r}$ coordinate and the time $t$ of the SSC,
which are the most important observational quantities, as discussed in the
previous paragraphs. This follows immediately from the relation between the
comoving coordinate $\chi$ and the conformal time $\eta$ found in Eqs.
(\ref{eqn3.15.2})-(\ref{eqn3.15.4}) for the propagation of a light signal.
This relation can be re-written as%

\begin{equation}
\chi=\sqrt{\left\vert k\right\vert }c(t_{0}-t)=\eta_{0}-\eta, \label{eqn4.24}%
\end{equation}
assuming that the emission of the light signal happens at coordinates
$(\chi,\eta)$ and the signal is received at $(\chi=0,\eta_{0})$. The relation
between $\mathbf{r}$ and the look-back time $t_{0}-t$ is therefore:%

\begin{equation}
\mathbf{r}=S_{\mathbf{k}}(\chi)=\left\{
\begin{array}
[c]{cc}%
\sin\left[  \sqrt{\left\vert k\right\vert }c(t_{0}-t)\right]  \ ; &
\mathbf{k}=+1\\
c(t_{0}-t)\ ; & \mathbf{k}=0\\
\sinh\left[  \sqrt{\left\vert k\right\vert }c(t_{0}-t)\right]  \ ; &
\mathbf{k}=-1
\end{array}
\right\}  . \label{eqn4.25}%
\end{equation}

These relations represent a simple and precise connection between the two
coordinates for the motion of a light signal over cosmological distances, but
with time measured by our current standard units.

\subsection{\label{sect:analysis_discussion}Analysis of the solutions for the
cosmic scale factor}

Equations (\ref{eqn4.4}), (\ref{eqn4.11}), (\ref{eqn4.17}), (\ref{eqn4.20}),
and (\ref{eqn4.23}) are our main result. They represent the closed-form
expressions for the cosmic scale factor $R$, or rather its ratio to the
present value, as a function of the \textquotedblleft look-back
time\textquotedblright\ or \textquotedblleft look-back
distance\textquotedblright\ in both SSC or RW coordinates. In the way they
were derived such equations are also valid for space-time coordinates in the
\textquotedblleft future,\textquotedblright\ so they represent the overall
evolution of the Universe.

We recall once again that, of the two parameters in our fundamental equations,
$\gamma$ is approximately determined by Mannheim's fits as a small positive
quantity. The other constant $k=-\frac{\gamma^{2}}{4}-\kappa$ is still
undetermined, in both magnitude and sign, since it includes the still unknown
$\kappa$ quantity. We will not assume $\kappa=0$, as in Mannheim's cosmology
\cite{Mannheim:2005bf}, since we note that, in the particular case of
$\kappa=0$, $k=-\frac{\gamma^{2}}{4}<0$, our solution in Eq. (\ref{eqn4.17})
would yield a very simple (and unlikely) exponential solution%

\begin{equation}
1+z=\frac{R(t_{0})}{R(t)}=\ \cosh[\frac{\gamma}{2}x]-\sinh[\frac{\gamma}%
{2}x]=e^{-\frac{\gamma}{2}c(t_{0}-t)}, \label{eqn4.26}%
\end{equation}
reminiscent of the classic steady state cosmology,\ a theory that we don't
believe can represent the physical reality of our Universe. We need therefore
to study in detail our fundamental solutions and obtain more accurate values
of the parameters from experimental observations \cite{Varieschi:2008va}.

\begin{figure}
[ptb]
\begin{center}
\fbox{\ifcase\msipdfoutput
\includegraphics[
height=5.5607in,
width=6.301in
]%
{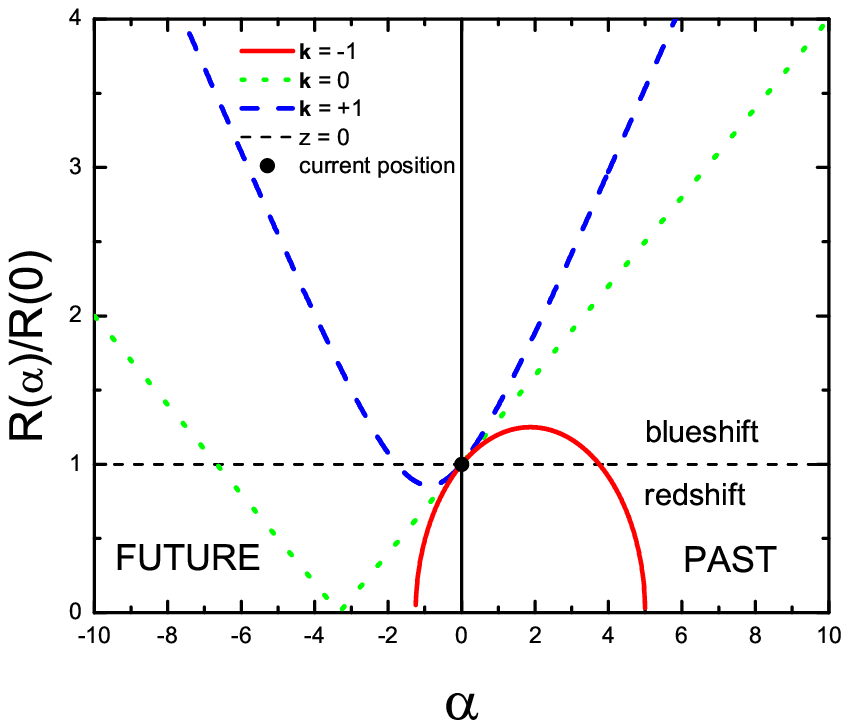}%
\else
\includegraphics[
height=5.5607in,
width=6.301in
]%
{C:/swp55/Docs/KINEMATICAL1/revised/revised2/revised3/arxiv_v2/graphics/figure1__1.pdf}%
\fi
}\caption[R functions in Eq. (\ref{eqn4.27}) are shown here for different
values of $\mathbf{k}$.]{R functions in Eq. (\ref{eqn4.27}) are shown here for
different values of $\mathbf{k}$: $\mathbf{k}=-1$ in red (solid),
$\mathbf{k}=0$ in green (dotted), and $\mathbf{k}=+1$ in blue (dashed), and
for a positive value of the parameter $\delta\simeq0.6$ (an unrealistically
large value; our current value $\delta=\delta(t_{0})$ will be shown to be
positive and close to zero \cite{Varieschi:2008va}).}%
\label{fig1}%
\end{center}
\end{figure}

To analyze the general form of our solutions it is better to plot the ratio
$R/R_{0}$, which shows directly the past (and future) evolution of the cosmic
scale factor. We also notice that all formulas can be rewritten using only
dimensionless quantities which are particularly convenient. Eq. (\ref{eqn4.4})
can be written as:%

\begin{align}
\frac{R(\alpha)}{R(0)}  &  =\sqrt{1+\delta\alpha+\frac{1}{4}(\delta
^{2}+\mathbf{k})\alpha^{2}}\label{eqn4.27}\\
\delta &  \equiv\frac{\gamma}{2\sqrt{\left\vert k\right\vert }};\ \alpha
\equiv2\sqrt{\left\vert k\right\vert }r\ ;\ \mathbf{k}=\frac{k}{\left\vert
k\right\vert }=0,\pm1\nonumber
\end{align}
where $\alpha$ becomes a dimensionless coordinate and the parameter $\delta$
is thought of as a possible time varying quantity. In Fig. \ref{fig1} we plot
these solutions, for the three values of $\mathbf{k}$, assuming a positive
current value $0<\delta=\delta(t_{0})<1$.%

We recall that the positive horizontal semi-axis is actually representing a
\textquotedblleft look-back\textquotedblright\ quantity, here expressed using
the dimensionless $\alpha\equiv2\sqrt{\left\vert k\right\vert }r$; on the
contrary the negative horizontal semi-axis represents \textquotedblleft
future\textquotedblright\ values of this variable, so that the universal
evolution of the cosmic scale factor should be observed following our curves
from right to left. The dot on the vertical axis represents our
\textquotedblleft current\textquotedblright\ position in the universal
evolution. We can clearly see that the only solution which indicates a
redshift in the past (values below the black thin dashed line for positive
$\alpha$) is the one for $\mathbf{k}=-1$ (red-solid, in Fig. \ref{fig1}). We
will therefore consider this particular solution as our fundamental candidate
to represent the evolution of the Universe.

The analytic characteristics of this solution (for $\mathbf{k}=-1$) are easily
determined for a given value of $\delta$ ($-1<\delta<1$): the zeroes of this
function are obtained for $\alpha=\frac{2}{1-\delta};-\frac{2}{1+\delta}$ or
for $r=1/(\sqrt{\left\vert k\right\vert }-\frac{\gamma}{2});-1/(\sqrt
{\left\vert k\right\vert }+\frac{\gamma}{2})$, respectively in the
\textquotedblleft past\textquotedblright\ and in the \textquotedblleft
future.\textquotedblright\ As discussed at the end of the previous section,
the variable $r$ (or $\alpha$) does not represent a cosmological distance, but
rather a simple coordinate in the Static Standard frame of reference,
therefore the values given above might represent initial or final
\textquotedblleft singularities\textquotedblright\ of the Universe, but only
if we were to measure the Universe with our fixed meter stick. It seems more
appropriate to consider them just as limiting values for our $r$ coordinate.

As already noted before, the region in the past closer to our current
location, for $0<\alpha<\alpha_{rs}=\frac{4\delta}{1-\delta^{2}}$, or
$0<r<r_{rs}=\gamma/(\left\vert k\right\vert -\frac{\gamma^{2}}{4})$, would
yield a blueshift. The redshift region occurs for $\alpha_{rs}=\frac{4\delta
}{1-\delta^{2}}<\alpha<\frac{2}{1-\delta}$ or $r_{rs}=\gamma/(\left\vert
k\right\vert -\frac{\gamma^{2}}{4})<r<1/(\sqrt{\left\vert k\right\vert }%
-\frac{\gamma}{2})$. The blueshift region (greatly exaggerated in Fig.
\ref{fig1}), could be just a small region around us, of the size of our Solar
System or part of our galaxy, depending on the current values of the
parameters $\gamma$ and $k$. The red curve in Fig. \ref{fig1} is also
obviously symmetric around its point of maximum, located at $\alpha_{\max
}=\frac{2\delta}{1-\delta^{2}}=\frac{\alpha_{rs}}{2}$, which corresponds to a
maximum blueshift $(R(\alpha)/R(0))_{\max}=1/\sqrt{1-\delta^{2}}$ or $z_{\min
}=\sqrt{1-\delta^{2}}-1$, the minimum value of $z$ in the blueshift region
(negative value).

In the previous paragraphs we started using dimensionless quantities, which
greatly simplify the analysis of our solutions. In particular, the
dimensionless $\delta\equiv\frac{\gamma}{2\sqrt{\left\vert k\right\vert }}$,
which enters all our fundamental equations, seems to be more important than
the single dimensionful parameters $\gamma$ and $k$. Plotting the solid red
curve of Fig. \ref{fig1}, for values of $\delta$ varying from $-1$ to $+1$,
would show a family of curves of similar shape and interpretation, just with
the point representing our \textquotedblleft current\textquotedblright%
\ position ($R(r)/R(0)=1$) shifting along the red curve from the limiting
position on the right of the graph (for $\delta\rightarrow-1$) to the limiting
position on the left (for $\delta\rightarrow+1$). Similar considerations would
also apply for the other two curves in the figure.

We recall that in modern cosmology a cosmic standard time coordinate should be
related to some property of the evolving Universe itself. Quoting from
Weinberg (\cite{Weinberg}, page 409): \textquotedblleft... several cosmic
scalar fields... are everywhere decreasing monotonically; choose any one of
these, say a scalar $S$, and let the time of any event be any definite
decreasing function $t(S)$ of the chosen scalar, when and where the event
occurs.\textquotedblright\ In view of the preceding discussion, it seems
possible that the role of the universal quantity $S$, only increasing rather
than decreasing, might be taken by $\delta\equiv\gamma/2\sqrt{\left\vert
k\right\vert }$, varying continuously from $-1$ to $+1$. Plotting our
fundamental solution ($\mathbf{k}=-1$) for $\delta$ increasing monotonically
from $-1$ to $1$, we would observe at first (for values of $\delta$ close to
$-1)$ a plot similar to the one of Fig. \ref{fig1}, but with the limiting
position on the right very close to the origin and a very steep slope of the
initial part of the plot: this is equivalent to a very fast initial expansion
of the Universe, a sort of \textquotedblleft inflationary\textquotedblright\ situation.

As $\delta$ increases in the negative interval $-1<\delta<0$, the expansion of
the Universe would seem to slow down: the limiting position in the past would
shift to the right in the plot, the red curve would \textquotedblleft
slide\textquotedblright\ to the right, subject to the condition of always
intersecting the vertical axis at $R(r)/R(0)=1$, and the slope of the curve at
the origin, corresponding to the expansion rate, would continuously decrease.
Similar behavior would be followed also by the $\mathbf{k}=0,+1$ solutions in
Fig. \ref{fig1}, but we consider these solutions not to be physically
relevant, therefore we concentrate our attention on the $\mathbf{k}=-1$
solution only (red-solid in all our figures). For $\delta=0$ the fundamental
($\mathbf{k}=-1$) solution would reduce to:%

\begin{equation}
\frac{R(t)}{R(t_{0})}=\sqrt{1-\frac{1}{4}\alpha^{2}}, \label{eqn4.28}%
\end{equation}
and the red curve in Fig. \ref{fig1} would simply be shifted in a symmetric
position with the maximum at $\alpha=0$, signaling that the maximum possible
expansion of the Universe has been reached (for an observer at the origin) and
that the expansion rate at the origin would now be zero. The zeroes of the
function would appear to be at $\alpha=\pm2$, with the Universe half way
through its cosmic evolution. The subsequent evolution would be seen by
letting the parameter $\delta$ run over positive values from $0$ to $1$. This
situation is again what is depicted in Fig. \ref{fig1}, corresponding to a
current positive value of $\delta$ (the plot in Fig. \ref{fig1} is actually
for $\delta\simeq0.6$, therefore greatly exaggerating the blueshift portion).

Our current value of $\delta=\delta(t_{0})$, that we will estimate in our
second paper \cite{Varieschi:2008va}, is probably positive and small,
therefore determining already a local contraction of the Universe (blueshift)
in a limited region around us, but still showing the past expansion (redshift)
in most of our visible Universe. For increasing positive values of $\delta$,
approaching the final $+1$ value, the rate of universal contraction would
increase, leading to the final situation in a totally symmetric way, compared
to the initial expansion.

The role of \textquotedblleft universal time\textquotedblright\ given to the
dimensionless parameter $\delta$, naturally prompts us to plot all our
fundamental solutions in terms of this universal quantity. This can be done
using Eq. (\ref{eqn4.27}) and by noting that the maximum of the red plot in
Fig. \ref{fig1} corresponds to a \textquotedblleft universal
time\textquotedblright\ $\delta=0$, in the sense that an observer placed at
that position, at the time of emission of the light which reaches us with the
maximum possible blueshift, would measure $\delta=0$ as his/her current
cosmological time.

\begin{figure}
[ptb]
\begin{center}
\fbox{\ifcase\msipdfoutput
\includegraphics[
height=4.5455in,
width=6.301in
]%
{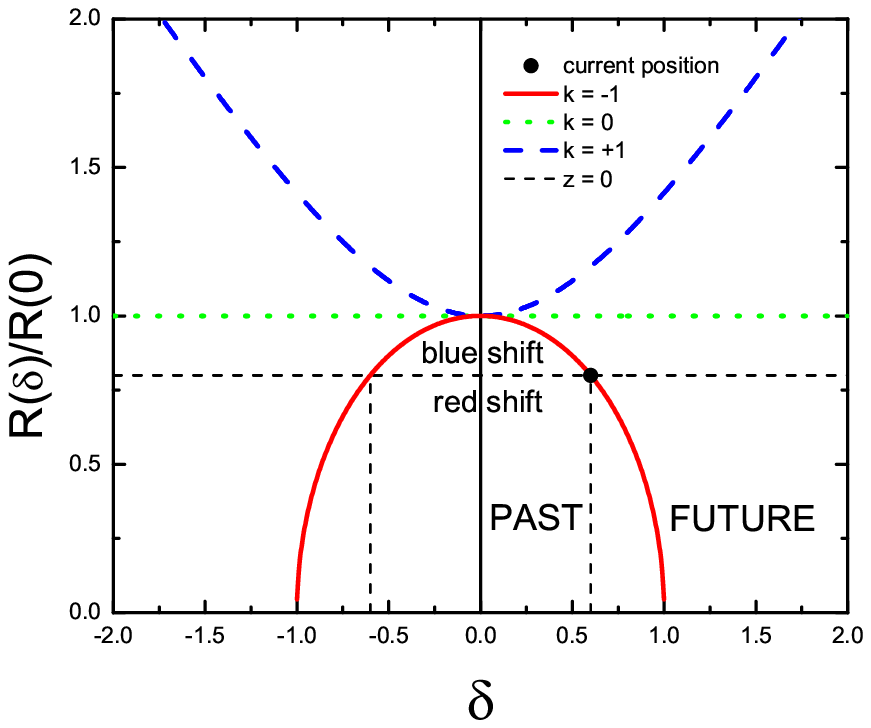}%
\else
\includegraphics[
height=4.5455in,
width=6.301in
]%
{C:/swp55/Docs/KINEMATICAL1/revised/revised2/revised3/arxiv_v2/graphics/figure2__2.pdf}%
\fi
}\caption[R functions in Eqs. (\ref{eqn4.29}), (\ref{eqn4.30}) are shown here
for different values of $\mathbf{k}$.]{R functions in Eqs. (\ref{eqn4.29}),
(\ref{eqn4.30}) are shown here for different values of $\mathbf{k}$:
$\mathbf{k}=-1$ in red (solid), $\mathbf{k}=0$ in green (dotted), and
$\mathbf{k}=+1$ in blue (dashed). The \textquotedblleft cosmological
time\textquotedblright\ $\delta$ is increasing from $-1$ to $+1$, so that the
evolution of the Universe proceeds from left to right along the plotted
curves. The present cosmological time is indicated as $\delta\simeq0.6$, an
unrealistically large value. Our current value of $\delta=\delta(t_{0})$ will
be shown to be positive and close to zero \cite{Varieschi:2008va}.}%
\label{fig2}%
\end{center}
\end{figure}

This maximum value occurs at $\alpha_{\max}=\frac{2\delta}{1-\delta^{2}}$ (or
for $r_{\max}=\gamma/2(\left\vert k\right\vert -\frac{\gamma^{2}}{4})$) and
the corresponding value is $R(\alpha_{\max})=R(\alpha=0)\frac{1}%
{\sqrt{1-\delta^{2}}}=R(\delta=0)$. But $R(\alpha=0)$ corresponds to the $R$
factor evaluated at the current value of the parameter $\delta$, i.e.,
$R(\delta)=R(\alpha=0)$, thus obtaining:\footnote{This argument is similar to
the assumption we made earlier, when converting the scale factor from being a
function of radial coordinates to a function of time. If $\delta$ plays the
role of a universal time, the Cosmological Principle would suggest that, for a
given value of $\delta$, all the locations in space should be equivalent,
therefore the scale factor $R$ should be properly a function just of the
cosmological time $\delta$.}%

\begin{equation}
R(\delta)=R(\delta=0)\sqrt{1-\delta^{2}};\ \mathbf{k}=-1, \label{eqn4.29}%
\end{equation}
whose plot as a function of the \textquotedblleft universal
time\textquotedblright\ $\delta$ is simply a semicircle of radius
corresponding to the maximum \textquotedblleft size\textquotedblright\ of the
Universe $R(\delta=0)$. Similar analysis would hold for the $\mathbf{k}=0,+1$
solutions, giving respectively:%

\begin{align}
R(\delta)  &  =R(\delta=0)\sqrt{1+\delta^{2}};\ \mathbf{k}=+1\label{eqn4.30}\\
R(\delta)  &  =R(\delta=0);\ \mathbf{k}=0.\nonumber
\end{align}
The three possible solutions for the evolution of the Universe, as a function
of $\delta$, are therefore remarkably simple, when plotted in these new
coordinate and are summarized in Fig. \ref{fig2}.%

In this figure, the \textquotedblleft universal time\textquotedblright%
\ $\delta$ now runs from \textquotedblleft past\textquotedblright\ to
\textquotedblleft future\textquotedblright\ for increasing values. The
red-solid plot shows our currently favorite cosmology of simple
\textquotedblleft semi-circular\textquotedblright\ evolution. The current
value of $\delta=\delta(t_{0})$ should be positive, such as the value
represented by the dot in the figure. Regions in the past below the horizontal
black dashed line represent the observed redshift, while the region above the
same horizontal thin black line indicates the local blueshift region. Again,
the $\delta(t_{0})$ value of the figure ($\delta(t_{0})\simeq0.6$) is greatly
exaggerated. The actual value should be very small and positive
\cite{Varieschi:2008va}.

The green-dotted $\mathbf{k}=0$ and the blue-dashed $\mathbf{k}=+1$ solutions
would yield respectively a static, constant size Universe or a parabolic,
first-contracting then-expanding Universe. These are not favored by
observation, in particular, for $\mathbf{k}=+1$, we would observe a redshift
at \textquotedblleft close\textquotedblright\ distances and then a blueshift
at \textquotedblleft large\textquotedblright\ cosmological distances.
Therefore our choice of a $\mathbf{k}=-1$ cosmology seems to be confirmed.

Using Eq. (\ref{eqn4.29}) with our current value of $\delta=\delta(t_{0})$,
i.e., $R\left[  \delta(t_{0})\right]  =R(\delta=0)\sqrt{1-\delta^{2}(t_{0})}$
and dividing through the same equation for an arbitrary value of $\delta$, we obtain%

\begin{equation}
1+z=\frac{R\left[  \delta(t_{0})\right]  }{R(\delta)}=\sqrt{\frac{1-\delta
^{2}(t_{0})}{1-\delta^{2}}}, \label{eqn4.30.1}%
\end{equation}
from which we can derive the most general connection between $\delta(t_{0})$,
$\delta$ and $z$ (for the case $\mathbf{k}=-1$):%

\begin{align}
z  &  =\sqrt{\frac{1-\delta^{2}(t_{0})}{1-\delta^{2}}}-1\label{eqn4.30.2}\\
\delta &  =\pm\frac{\sqrt{\delta^{2}(t_{0})+z(z+2)}}{(1+z)}.\nonumber
\end{align}
Alternatively, we can combine Eq. (\ref{eqn4.30.1}) with Eq. (\ref{eqn4.27}),
for the case $\mathbf{k}=-1$, namely $\sqrt{1-\delta^{2}}/\sqrt{1-\delta
^{2}(t_{0})}=\sqrt{1+\delta(t_{0})\alpha+\frac{1}{4}\left[  \delta^{2}%
(t_{0})-1\right]  \alpha^{2}}$, and solve it for $\alpha$ and $\delta$:%

\begin{align}
\alpha &  =2\frac{\delta(t_{0})-\delta}{1-\delta^{2}(t_{0})} \label{eqn4.30.3}%
\\
\delta &  =\delta(t_{0})-\frac{1}{2}\left[  1-\delta^{2}(t_{0})\right]
\alpha.\nonumber
\end{align}

\subsection{\label{sect:other_plots}The other fundamental solutions}

\begin{figure}
[ptb]
\begin{center}
\fbox{\ifcase\msipdfoutput
\includegraphics[
height=5.3575in,
width=6.301in
]%
{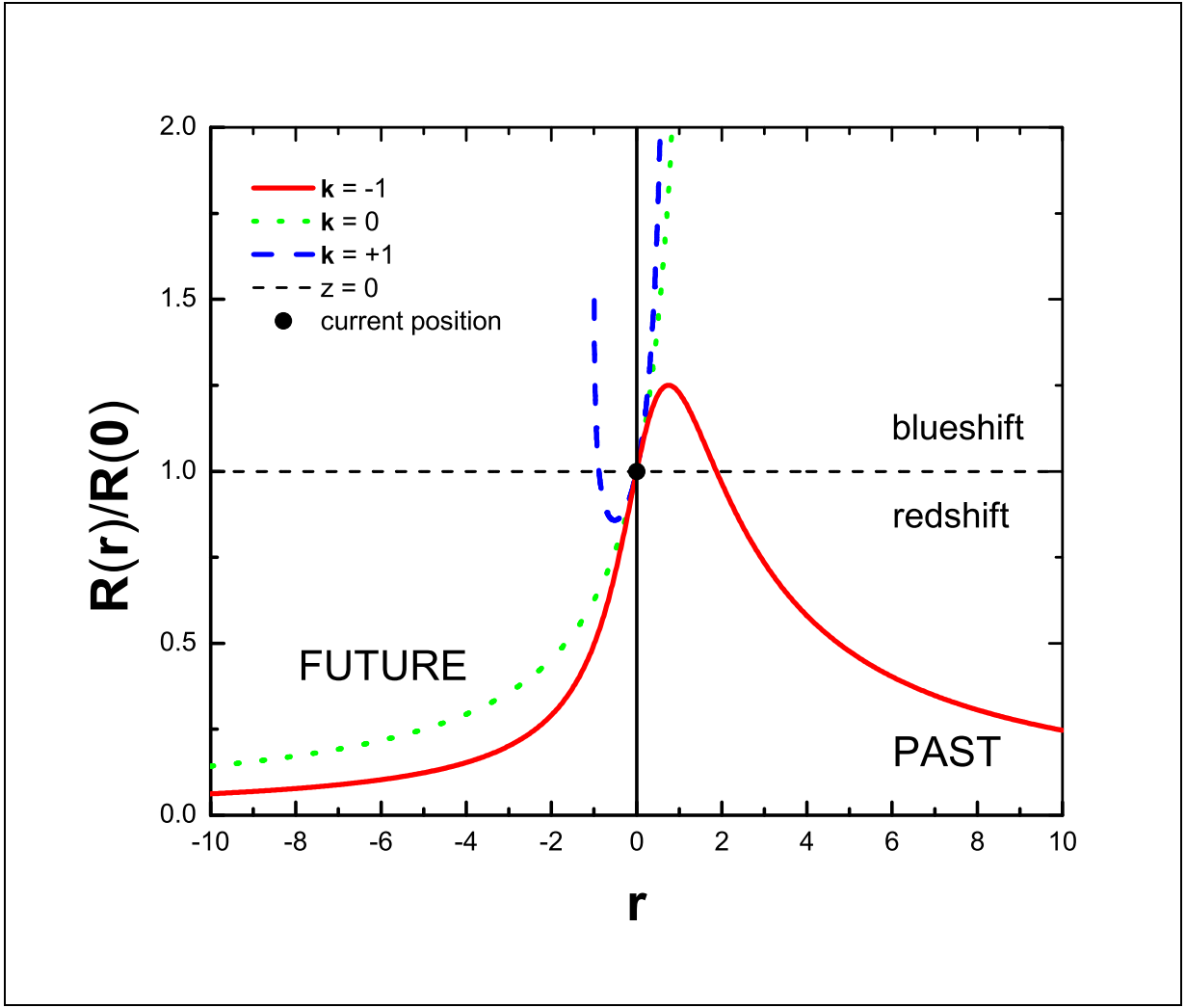}%
\else
\includegraphics[
height=5.3575in,
width=6.301in
]%
{C:/swp55/Docs/KINEMATICAL1/revised/revised2/revised3/arxiv_v2/graphics/figure3__3.pdf}%
\fi
}\caption[R functions in Eq. (\ref{eqn4.31}) are shown for different values of
$\mathbf{k}$.]{R functions in Eq. (\ref{eqn4.31}) are shown for different
values of $\mathbf{k}$: $\mathbf{k}=-1$ in red (solid), $\mathbf{k}=0$ in
green (dotted), and $\mathbf{k}=+1$ in blue (dashed). Again, we used an
unrealistic $\delta\simeq0.6$ to make the illustration more legible.}%
\label{fig3}%
\end{center}
\end{figure}

Most of the analysis in the previous section can be repeated also for the
other fundamental solutions. From Eq. (\ref{eqn4.11}) we obtain:%

\begin{equation}
\frac{\mathbf{R(r)}}{\mathbf{R(0)}}=\frac{1}{\sqrt{1-\mathbf{k\ r}^{2}}%
-\delta\mathbf{r}}\mathbf{\ ;\ \ k}=0,\pm1, \label{eqn4.31}%
\end{equation}
as a function of dimensionless quantities (we recall that $\mathbf{r}$ can be
considered dimensionless, following the note before Eq. (\ref{eqn3.15})). In
Fig. \ref{fig3} we plot these solutions for a positive value of $\delta$. The
same comments of Fig. \ref{fig1} are applicable here, only the shape of the
curves is different due to the transformation between coordinates $r$ and
$\mathbf{r}$, described in Eq. (\ref{eqn4.9}). The $\mathbf{k}=-1$ solution,
red-solid in the figure, is still the only cosmologically viable.%

Again, the cosmological evolution from past to future is seen following the
red curve from right to left. The past redshift region is followed by a local
blueshift region. The black dot indicates the present situation, but we still
used an unrealistic value $\delta\simeq0.6$ to plot the curves. In the actual
situation \cite{Varieschi:2008va}, for a small positive $\delta$, we would be
placed near the expansion maximum, at the beginning of the \textquotedblleft
contraction\textquotedblright\ phase. The main difference between Fig.
\ref{fig1} and Fig. \ref{fig3} is that the red continuous curve, as a function
of $\mathbf{r}$, does not show any initial or final \textquotedblleft
singularity,\textquotedblright\ i.e., the evolution function only approaches
zero for $\mathbf{r}\rightarrow\pm\infty$. Since astronomical distances, such
as the luminosity distance or others, are defined using $\mathbf{r}$, the
Universe will not show any initial or final \textquotedblleft
singularity\textquotedblright\ when measured using these coordinates. On the
contrary, such initial and final points were shown to be present in Fig.
\ref{fig1} or Fig. \ref{fig2}, where the \textquotedblleft universal
time\textquotedblright\ $\delta$ was used.

In other words, the description of the Universe through the parameter $\delta$
would suggest the existence of an \textquotedblleft initial\textquotedblright%
\ and a \textquotedblleft final\textquotedblright\ singularity, thus not
contradicting our standard \textquotedblleft Big Bang\textquotedblright%
\ intuition, but tracing the origin of the Universe with the $\mathbf{r}$
coordinate would not describe the initial singularity in terms of a finite
\textquotedblleft distance\textquotedblright\ corresponding to a zero of the
$\mathbf{R}$ function. A similar \textquotedblleft dual\textquotedblright%
\ interpretation will also be found for the time coordinates in the following
paragraphs, although with an inverted role played by the coordinates $t$ and
$\mathbf{t}$.

As for the other analytical properties of the $\mathbf{k}=-1$ solution in Fig.
\ref{fig3}, the positions of zero redshift are located at the origin
$\mathbf{r}=0$ and at $\mathbf{r}_{rs}=\frac{2\delta}{1-\delta^{2}}$. The
position of maximum expansion is $\mathbf{r}_{\max}=\frac{\delta}%
{\sqrt{1-\delta^{2}}}$, corresponding to a maximum blueshift
$(\mathbf{R(r)/R(0)})_{\max}=1/\sqrt{1-\delta^{2}}$ or $z_{\min}%
=\sqrt{1-\delta^{2}}-1$, as already found with the graph in Fig. \ref{fig1}.
The $\mathbf{R(r)}$ curve is not symmetric around its point of maximum, as it
was for the curve in $r$, due to the transformation between these two
coordinates. The points of inflection of the red curve in Fig. \ref{fig3},
representing the position of change between a positive and negative
acceleration of the expansion, can also be easily found analytically from Eq.
(\ref{eqn4.31}), but their expression is rather cumbersome and will be omitted here.

It is possible to repeat with the $\mathbf{r}$ coordinate the same reasoning
done with $r$, to obtain the solution in terms of the universal parameter
$\delta$, by using the fact that the maximum expansion position $\mathbf{r}%
_{\max}=\frac{\delta}{\sqrt{1-\delta^{2}}}$ (for $\delta=\delta(\mathbf{t}%
_{0})$ our local current value) is assumed to correspond to $\delta
(\mathbf{r}_{\max})=0$, when this parameter is measured at the position of
maximum expansion. In this way we find the same solutions already expressed in
Eqs. (\ref{eqn4.29}), (\ref{eqn4.30}) and plotted in Fig. \ref{fig2}.

\begin{figure}
[ptb]
\begin{center}
\fbox{\ifcase\msipdfoutput
\includegraphics[
height=5.5218in,
width=6.301in
]%
{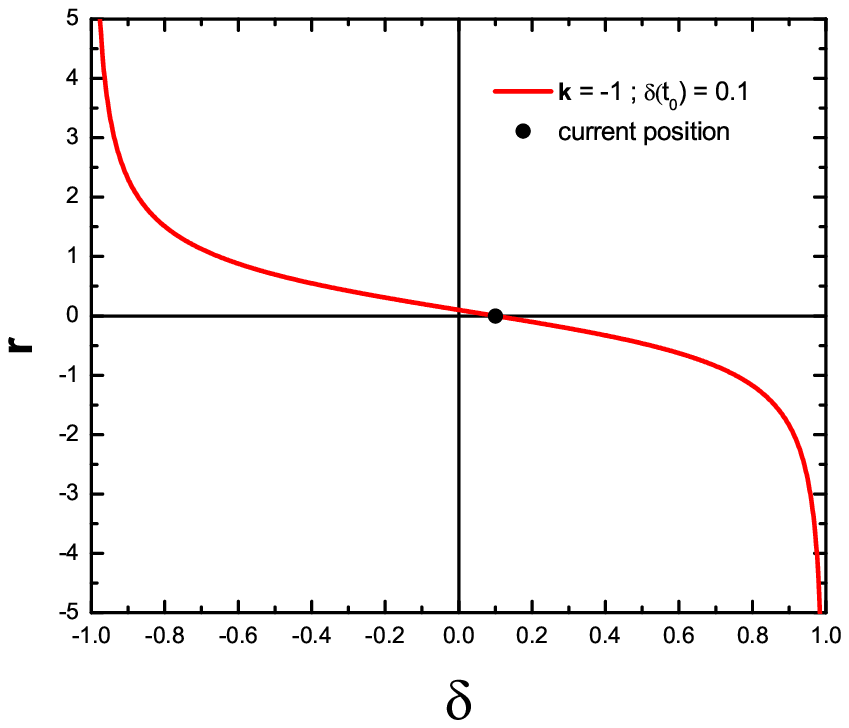}%
\else
\includegraphics[
height=5.5218in,
width=6.301in
]%
{C:/swp55/Docs/KINEMATICAL1/revised/revised2/revised3/arxiv_v2/graphics/figure4__4.pdf}%
\fi
}\caption[The connection between $\mathbf{r}$ and $\delta$ is illustrated
here.]{The connection between $\mathbf{r}$ and $\delta$ is illustrated here
for a value $\delta(\mathbf{t}_{0})\simeq0.1$. The correct value of the
cosmological time will be determined later \cite{Varieschi:2008va}.}%
\label{fig4}%
\end{center}
\end{figure}

In the same way, as it was done for the variable $\alpha$, we can obtain the
connections between $\mathbf{r}$, $\delta$ and our current value
$\delta(\mathbf{t}_{0})$ (for the $\mathbf{k}=-1$ case):%

\begin{align}
\mathbf{r}  &  =\frac{\delta(\mathbf{t}_{0})-\delta}{\sqrt{\left[
1-\mathbf{\delta}^{2}(\mathbf{t}_{0})\right]  \left[  1-\mathbf{\delta}%
^{2}\right]  }}\label{eqn4.32}\\
\delta &  =\frac{\mathbf{r-}\delta(\mathbf{t}_{0})\sqrt{1+\mathbf{r}^{2}}%
}{\delta(\mathbf{t}_{0})\mathbf{r}-\sqrt{1+\mathbf{r}^{2}}}.\nonumber
\end{align}
Considering, for example, the current value as $\delta=\delta(\mathbf{t}%
_{0})\cong0.1$, the previous equation is plotted Fig. \ref{fig4}. We can see
from this figure that the initial and final values for the cosmological time,
$\delta=\mp1$, correspond to infinite values of the $\mathbf{r}$ coordinate,
as already remarked.%

It is also immediate to determine the relation between $\mathbf{r}$ and the
redshift parameter $z$. Directly from Eq. (\ref{eqn4.11}) for the case
$\mathbf{k}=-1$, we obtain such relation and its inverse:%

\begin{align}
z  &  =\sqrt{1+\mathbf{r}^{2}}-\delta\mathbf{r-}1\label{eqn4.35}\\
\mathbf{r}  &  =\frac{\delta(1+z)\pm\sqrt{\delta^{2}+z(z+2)}}{(1-\delta^{2}%
)}=\frac{\delta(1+z)\pm\sqrt{(1+z)^{2}-(1-\delta^{2})}}{(1-\delta^{2}%
)},\nonumber
\end{align}
where the inverse expression holds for $z\geq z_{\min}=\sqrt{1-\delta^{2}}-1$.

\begin{figure}
[ptb]
\begin{center}
\fbox{\ifcase\msipdfoutput
\includegraphics[
height=5.5607in,
width=6.301in
]%
{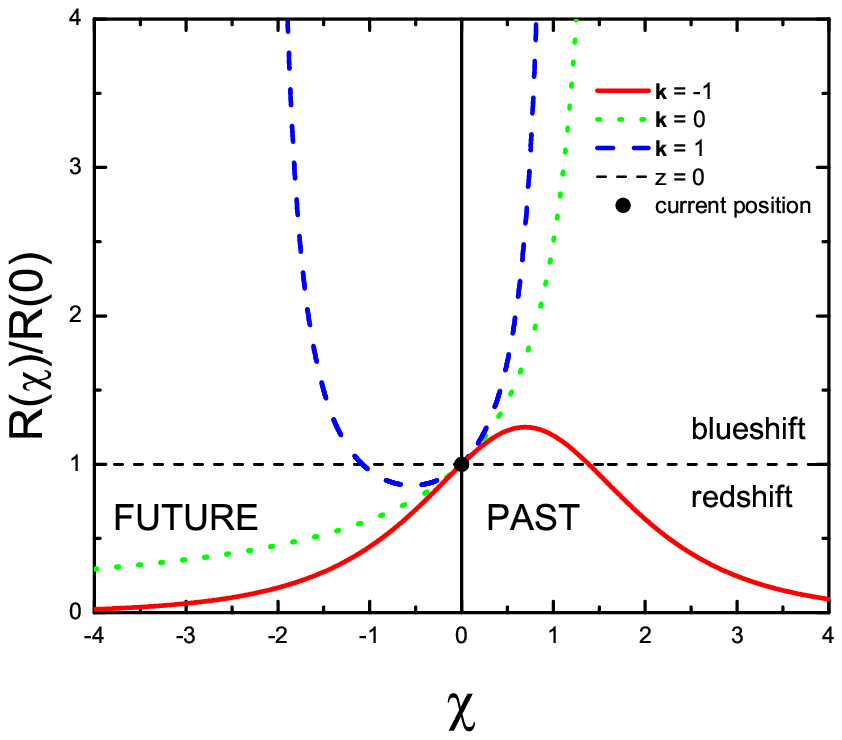}%
\else
\includegraphics[
height=5.5607in,
width=6.301in
]%
{C:/swp55/Docs/KINEMATICAL1/revised/revised2/revised3/arxiv_v2/graphics/figure5__5.pdf}%
\fi
}\caption[R functions in Eq. (\ref{eqn4.36}) are shown here for different
values of $\mathbf{k}$.]{R functions in Eq. (\ref{eqn4.36}) are shown here for
different values of $\mathbf{k}$: $\mathbf{k}=-1$ in red (solid),
$\mathbf{k}=0$ in green (dotted), and $\mathbf{k}=+1$ in blue (dashed). We use
again $\delta\simeq0.6$, as in previous illustrations.}%
\label{fig5}%
\end{center}
\end{figure}

Similar considerations apply to the time dependent solutions. Eq.
(\ref{eqn4.20}) can also be written in dimensionless form:%

\begin{align}
\frac{R(t)}{R(t_{0})}  &  =\left[  \cos\chi-\delta\sin\chi\right]
^{-1}\ ;\ k>0\label{eqn4.36}\\
\frac{R(t)}{R(t_{0})}  &  =\left[  1-\delta\chi\right]  ^{-1}%
\ ;\ k=0\nonumber\\
\frac{R(t)}{R(t_{0})}  &  =\left[  \cosh\chi-\delta\sinh\chi\right]
^{-1}\ ;\ k<0\nonumber\\
\chi &  \equiv\sqrt{\left\vert k\right\vert }x=\sqrt{\left\vert k\right\vert
}c(t_{0}-t)\nonumber
\end{align}
and is illustrated in Fig. \ref{fig5} (as usual for an unrealistically large
value of $\delta\simeq0.6$).%

We can find the main characteristics of the $\mathbf{k}=-1$ solution (in
red-solid) as we have done in the previous cases. We first remark that the
variable $\chi\equiv\sqrt{\left\vert k\right\vert }x=\sqrt{\left\vert
k\right\vert }c(t_{0}-t)$ is proportional to the time interval $(t_{0}-t)$ in
Static Standard Coordinates which, as discussed at the end of Sect.
\ref{sect:time_dependent_form}, is the time variable we will use in Cosmology
for time measurements, age determinations, etc., so the expressions in Eq.
(\ref{eqn4.36}) are actually more important than those in the RW time variable
that we will discuss later. The $\chi$ coordinate used here also corresponds
to the comoving coordinate of Eq. (\ref{eqn3.15.1}) as already discussed at
the end of Sect. \ref{sect:from_SSC_to_RW}.

As for the solution in $\mathbf{r}$, we note immediately that the red curve in
Fig. \ref{fig5} doesn't show initial or final singularities: the Universe
appears to be infinitely old and will never end if measured with our time
standard. We find that $\chi_{rs}=\operatorname{arccosh}\left[  (1+\delta
^{2})/(1-\delta^{2})\right]  =2\operatorname{arctanh}\delta$ or $(t_{0}%
-t)_{rs}=\frac{2}{\sqrt{\left\vert k\right\vert }c}\operatorname{arctanh}%
\delta$, for the look-back time at which redshift starts being observed. The
red curve has a maximum at $\chi_{\max}=\operatorname{arctanh}\delta$ or
$(t_{0}-t)_{\max}=\frac{1}{\sqrt{\left\vert k\right\vert }c}%
\operatorname{arctanh}\delta$ (again we find $(R(t)/R(0))_{\max}%
=1/\sqrt{1-\delta^{2}}$ or $z_{\min}=\sqrt{1-\delta^{2}}-1$) and it is
evidently symmetric around this point of maximum expansion of the Universe.

In fact, it is easy to check that with a translation of the $\chi$ coordinate,
$\chi=\chi_{\max}+\widetilde{\chi}$, which brings the origin of the new
$\widetilde{\chi}$ coordinate to the point of maximum, we have $\left[
\cosh\left(  \chi\right)  -\delta\sinh\left(  \chi\right)  \right]
=\sqrt{1-\delta^{2}}\cosh(\widetilde{\chi})$ so that:%

\begin{equation}
\frac{R(t)}{R(t_{0})}=\left[  \sqrt{1-\delta^{2}}\cosh(\widetilde{\chi
})\right]  ^{-1}=\frac{\operatorname{sech}(\widetilde{\chi})}{\sqrt
{1-\delta^{2}}}. \label{eqn4.37}%
\end{equation}
The time dependent evolution function becomes therefore a very simple function
when described in terms of the $\widetilde{\chi}$ coordinate and we can obtain
the $\delta$ dependent form of the evolution factor, namely $R(\delta
)=R(\delta=0)\sqrt{1-\delta^{2}}$ as in Eq. (\ref{eqn4.29}), by considering
that the maximum for $\chi_{\max}=\operatorname{arctanh}\delta$ corresponds to
$\delta=0$ and using this information inside Eq. (\ref{eqn4.37}).

\begin{figure}
[ptb]
\begin{center}
\fbox{\ifcase\msipdfoutput
\includegraphics[
height=5.354in,
width=6.301in
]%
{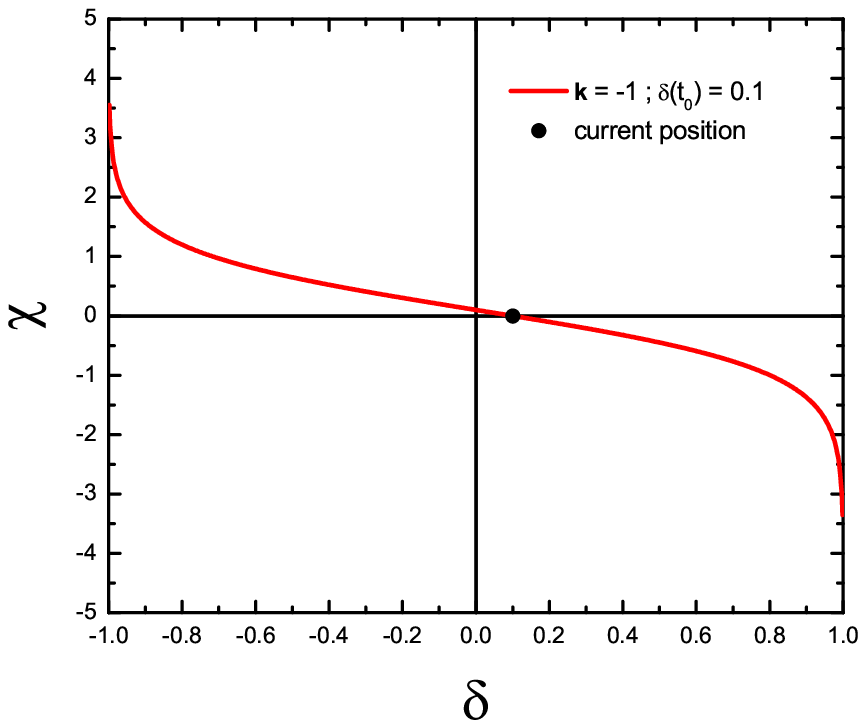}%
\else
\includegraphics[
height=5.354in,
width=6.301in
]%
{C:/swp55/Docs/KINEMATICAL1/revised/revised2/revised3/arxiv_v2/graphics/figure6__6.pdf}%
\fi
}\caption[The connection between $\chi$ and $\delta$ is illustrated here.]{The
connection between $\chi$ and $\delta$ is illustrated here for a value
$\delta(t_{0})\protect\cong0.1$. The correct value of the cosmological time
will be determined later \cite{Varieschi:2008va}.}%
\label{fig6}%
\end{center}
\end{figure}

This yields also to the general connection between $\delta$ and $\chi$, i.e.,
how the fundamental cosmological parameter $\delta$ changes with our time.
Following the same steps which led us to Eq. (\ref{eqn4.30.3}), we obtain for
$\mathbf{k}=-1$:%
\begin{align}
\chi &  =\operatorname{arctanh}\delta(t_{0})-\operatorname{arctanh}%
\delta\label{eqn4.38}\\
\delta &  =-\tanh[\chi-\operatorname{arctanh}\delta(t_{0})].\nonumber
\end{align}
To provide a practical example, if we assume $\delta=\delta(t_{0})\cong0.1$,
the previous equation is plotted in Fig. \ref{fig6}.%

The connection between $\chi$ (or the look-back time $t_{0}-t$) and $z$
immediately follows from Eqs. (\ref{eqn4.17}) or (\ref{eqn4.36}), for
$\mathbf{k}=-1$:%

\begin{align}
z  &  =\left[  \cosh\chi-\delta\sinh\chi\right]  -1\label{eqn4.40}\\
\chi &  =\operatorname{arcsinh}\frac{\delta(1+z)\pm\sqrt{\delta^{2}+z(z+2)}%
}{(1-\delta^{2})}\nonumber
\end{align}
From this expression we can also determine the following relation:%

\begin{equation}
\sinh\chi-\delta\cosh\chi=\pm\sqrt{\delta^{2}+z(z+2)}=\pm\sqrt{(1+z)^{2}%
-(1-\delta^{2})}, \label{eqn4.41}%
\end{equation}
which is useful to compute the time derivatives of the cosmic scale factor as
a function also of the redshift $z$:%

\begin{align}
\overset{\cdot}{R}(t)  &  =R(t_{0})\sqrt{\left\vert k\right\vert }%
c\frac{\left[  \sinh\chi-\delta\cosh\chi\right]  }{\left[  \cosh\chi
-\delta\sinh\chi\right]  ^{2}}=\pm R(t_{0})\sqrt{\left\vert k\right\vert
}c\frac{\sqrt{(1+z)^{2}-(1-\delta^{2})}}{(1+z)^{2}}\label{eqn4.42}\\
\overset{\cdot\cdot}{R}(t)  &  =-R(t_{0})\left\vert k\right\vert c^{2}%
\frac{\left[  \cosh\chi-\delta\sinh\chi\right]  ^{2}-2\left[  \sinh\chi
-\delta\cosh\chi\right]  ^{2}}{\left[  \cosh\chi-\delta\sinh\chi\right]  ^{3}%
}=\nonumber\\
&  =R(t_{0})\left\vert k\right\vert c^{2}\frac{(1+z)^{2}-2(1-\delta^{2}%
)}{(1+z)^{3}}.\nonumber
\end{align}

\begin{figure}
[ptb]
\begin{center}
\fbox{\ifcase\msipdfoutput
\includegraphics[
height=5.5607in,
width=6.301in
]%
{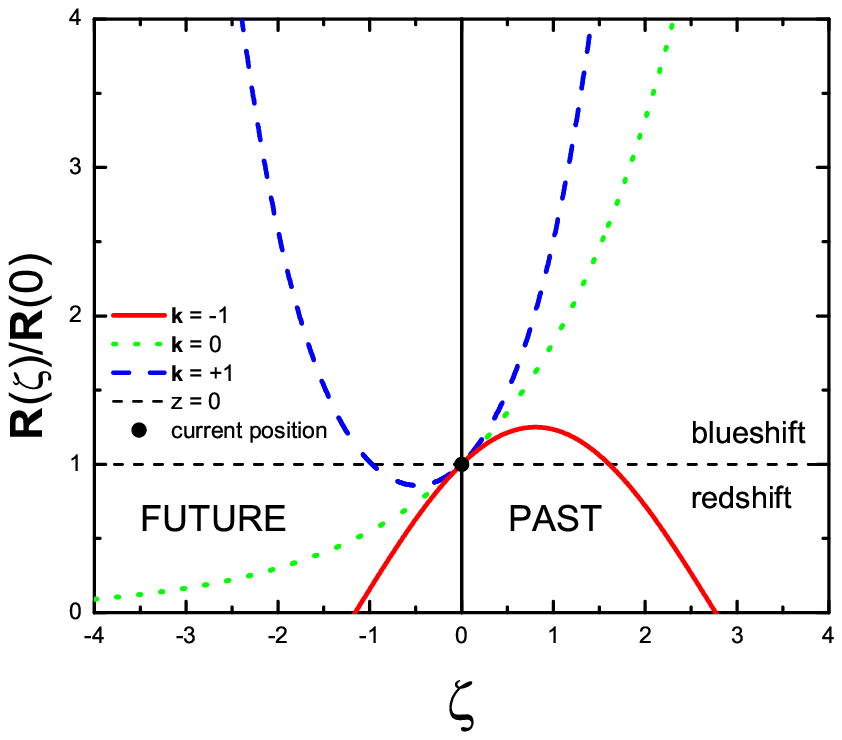}%
\else
\includegraphics[
height=5.5607in,
width=6.301in
]%
{C:/swp55/Docs/KINEMATICAL1/revised/revised2/revised3/arxiv_v2/graphics/figure7__7.pdf}%
\fi
}\caption[R functions in Eq. (\ref{eqn4.43}) are shown here for different
values of $\mathbf{k}$.]{R functions in Eq. (\ref{eqn4.43}) are shown here for
different values of $\mathbf{k}$: $\mathbf{k}=-1$ in red (solid),
$\mathbf{k}=0$ in green (dotted), and $\mathbf{k=+1}$ in blue (dashed). We use
again $\delta\simeq0.6$, as an example.}%
\label{fig7}%
\end{center}
\end{figure}

Finally, Eq. (\ref{eqn4.23}) is easily expressed as%

\begin{align}
\frac{\mathbf{R(t)}}{\mathbf{R(t}_{0}\mathbf{)}}  &  =\left[  \cosh\left(
\sqrt{1+\delta^{2}}\zeta\right)  +\frac{\delta}{\sqrt{1+\delta^{2}}}%
\sinh\left(  \sqrt{1+\delta^{2}}\zeta\right)  \right]  \ ;\ \mathbf{k}%
=1\label{eqn4.43}\\
\frac{\mathbf{R(t)}}{\mathbf{R(t}_{0}\mathbf{)}}  &  =e^{\delta\zeta
}\ ;\ \mathbf{k}=0\nonumber\\
\frac{\mathbf{R(t)}}{\mathbf{R(t}_{0}\mathbf{)}}  &  =\left[  \cos\left(
\sqrt{1-\delta^{2}}\zeta\right)  +\frac{\delta}{\sqrt{1-\delta^{2}}}%
\sin\left(  \sqrt{1-\delta^{2}}\zeta\right)  \right]  \ ;\ \mathbf{k}%
=-1\nonumber\\
\zeta &  \equiv\frac{\mathbf{x}}{\mathbf{R(t}_{0}\mathbf{)}}=\frac
{c(\mathbf{t}_{0}-\mathbf{t)}}{\mathbf{R(t}_{0}\mathbf{)}}\nonumber
\end{align}
and is illustrated in Fig. \ref{fig7} for $\delta\simeq0.6$. In this case, for
the $\mathbf{k}=-1$ solution, the $\zeta$ variable is limited within the
interval $\frac{1}{\sqrt{1-\delta^{2}}}\arctan\left(  -\frac{\sqrt
{1-\delta^{2}}}{\delta}\right)  <\zeta<\frac{1}{\sqrt{1-\delta^{2}}}\left[
\arctan\left(  -\frac{\sqrt{1-\delta^{2}}}{\delta}\right)  +\pi\right]  $,
$\zeta_{\max}=\frac{1}{\sqrt{1-\delta^{2}}}\arctan\left(  \frac{\delta}%
{\sqrt{1-\delta^{2}}}\right)  $ and $\zeta_{rs}=\frac{2}{\sqrt{1-\delta^{2}}%
}\arctan\left(  \frac{\delta}{\sqrt{1-\delta^{2}}}\right)  $.%

For all the solutions discussed in this section and described by Eqs.
(\ref{eqn4.31}), (\ref{eqn4.36}), (\ref{eqn4.43}), it is possible to express
them as a function of the \textquotedblleft universal time\textquotedblright%
\ $\delta$, as we have done for Eq. (\ref{eqn4.27}). In all cases the result
is obviously always the same, as expressed by Eqs. (\ref{eqn4.29}),
(\ref{eqn4.30}) and shown in Fig. \ref{fig2}. All our expressions of the
cosmic scale factor $R/R_{0}$ just differ in the coordinates used, but they
represent the same cosmological function.

In summary, our description of the past and future evolution of the Universe,
through the cosmic scale factor $R$, can be done by using any of the six
kinematical dimensionless variables $\alpha$, $\mathbf{r}$, $\chi$, $\zeta$,
$z$ and $\delta$. They are all related to one another by the transformations
outlined above. In Table 1 we have included all the possible connecting
formulas, for the $\mathbf{k}=-1$ case, adding those not explicitly analyzed
in the current section.

In this Table, as well as in all the equations above, special care is to be
given to the meaning of the universal time $\delta$. Whenever we connect
together any two of the dimensionless variables $\alpha$, $\mathbf{r}$, $\chi
$, $\zeta$, $z$, such as in Eqs. (\ref{eqn4.11}), (\ref{eqn4.27}),
(\ref{eqn4.35}), (\ref{eqn4.40}), these relations are assumed to hold for a
given, fixed value of the cosmological time. This value is simply indicated as
$\delta$ in all these formulas, but typically refers to the current value
$\delta=\delta(t_{0})$, which will be studied in our next paper
\cite{Varieschi:2008va} (or can be assumed to be in the range $-1<\delta<1$).

On the contrary, when we specify a direct connection between one of the
dimensionless variables $\alpha$, $\mathbf{r}$, $\chi$, $\zeta$, $z$ and the
cosmological time $\delta$, such as in Eqs. (\ref{eqn4.32}) or (\ref{eqn4.38}%
), we describe how this variable is changing together with $\delta$, as seen
from an observer at current time $\delta(t_{0})$, so that we have to specify
both quantities, $\delta$ and $\delta(t_{0})$, in these expressions.%

\begin{sidewaystable}[tbp] \centering
\begin{tabular}
[c]{||c|c|c|c||}\hline\hline
${\small \star}$ & \frame{$\alpha$} & \frame{$\mathbf{r}$} & \frame{$\chi$%
}\\\hline
\frame{$\alpha$} & ${\small \star}$ & ${\small \alpha=}\frac{2\mathbf{r}%
}{\sqrt{1+\mathbf{r}^{2}}-\delta\mathbf{r}}$ & ${\small \alpha=}\frac{2}%
{\coth\chi-\delta}$\\\hline
\frame{$\mathbf{r}$} & $\mathbf{r}{\small =}\frac{\operatorname{signum}\left(
\alpha\right)  }{\sqrt{\left(  \frac{2}{\alpha}+\delta\right)  ^{2}-1}}$ &
${\small \star}$ & $\mathbf{r}{\small =}\sinh${\small $\chi$}\\\hline
\frame{$\chi$} & {\small $\chi$}${\small =}\operatorname{arccoth}\left(
\frac{2}{\alpha}+\delta\right)  $ & {\small $\chi$}${\small =}%
\operatorname{arcsinh}\mathbf{r}$ & ${\small \star}$\\\hline
\frame{$\zeta$} & ${\small \zeta=}\frac{2}{\sqrt{1-\delta^{2}}}%
\operatorname{arccot}\left[  \frac{\frac{2}{\alpha}+\operatorname{signum}%
(\alpha)\sqrt{\left(  \frac{2}{\alpha}+\delta\right)  ^{2}-1}}{\sqrt
{1-\delta^{2}}}\right]  $ & ${\small \zeta=}\frac{2}{\sqrt{1-\delta^{2}}%
}\operatorname{arccot}\left(  \frac{1+\sqrt{1+\mathbf{r}^{2}}-\delta
\mathbf{r}}{\mathbf{r}\sqrt{1-\delta^{2}}}\right)  $ & ${\small \zeta=}%
\frac{2}{\sqrt{1-\delta^{2}}}\operatorname{arccot}\left(  \frac{\coth
\frac{\chi}{2}-\delta}{\sqrt{1-\delta^{2}}}\right)  $\\\hline
\frame{$z$} & ${\small z=}\frac{1}{\sqrt{1+\delta\alpha+\frac{1}{4}(\delta
^{2}-1)\alpha^{2}}}{\small -1}$ & ${\small z=}\sqrt{1+\mathbf{r}^{2}%
}{\small -\delta\mathbf{r}-1}$ & ${\small z=}\cosh${\small $\chi$%
}${\small -\delta}\sinh${\small $\chi$}${\small -1}$\\\hline
\frame{$\delta$} & ${\small \delta=}\delta(t_{0})-\frac{1}{2}\left[
1-\delta^{2}(t_{0})\right]  \alpha$ & ${\small \delta=}\frac{\mathbf{r-}%
\delta(\mathbf{t}_{0})\sqrt{1+\mathbf{r}^{2}}}{\delta(\mathbf{t}%
_{0})\mathbf{r-}\sqrt{1+\mathbf{r}^{2}}}$ & ${\small \delta=-}\tanh{\small [}%
${\small $\chi$}${\small -}\operatorname{arctanh}{\small \delta(t}%
_{0}{\small )]}$\\\hline
${\small \star}$ & \frame{$\zeta$} & \frame{$z$} & \frame{$\delta$}\\\hline
\frame{$\alpha$} & ${\small \alpha=}\frac{4\left[  \sqrt{1-\delta^{2}}%
\cot\left(  \sqrt{1-\delta^{2}}\frac{\zeta}{2}\right)  +\delta\right]
\sin^{2}\left(  \sqrt{1-\delta^{2}}\frac{\zeta}{2}\right)  }{\left(
1-\delta^{2}\right)  }$ & ${\small \alpha=2}\frac{\delta(1+z)\pm\sqrt
{\delta^{2}+z(z+2)}}{(1-\delta^{2})(1+z)}$ & ${\small \alpha=}2\frac
{\delta(t_{0})-\delta}{1-\delta^{2}(t_{0})}$\\\hline
\frame{$\mathbf{r}$} & $\mathbf{r}{\small =2}\frac{\left[  \sqrt{1-\delta^{2}%
}\cot\left(  \sqrt{1-\delta^{2}}\frac{\zeta}{2}\right)  +\delta\right]
}{\left[  \sqrt{1-\delta^{2}}\cot\left(  \sqrt{1-\delta^{2}}\frac{\zeta}%
{2}\right)  +\delta\right]  ^{2}-1}$ & $\mathbf{r}{\small =}\frac
{\delta(1+z)\pm\sqrt{\delta^{2}+z(z+2)}}{(1-\delta^{2})}$ & $\mathbf{r}%
{\small =}\frac{\delta(\mathbf{t}_{0})-\delta}{\sqrt{1-\delta^{2}%
(\mathbf{t}_{0})}\sqrt{1-\delta^{2}}}$\\\hline
\frame{$\chi$} & {\small $\chi$}${\small =2}\operatorname{arccoth}\left[
\sqrt{1-\delta^{2}}\cot\left(  \sqrt{1-\delta^{2}}\frac{\zeta}{2}\right)
+\delta\right]  $ & {\small $\chi$}${\small =}\operatorname{arcsinh}\left[
\frac{\delta(1+z)\pm\sqrt{\delta^{2}+z(z+2)}}{(1-\delta^{2})}\right]  $ &
{\small $\chi$}${\small =}\operatorname{arctanh}{\small \delta(t}%
_{0}{\small )-}\operatorname{arctanh}{\small \delta}$\\\hline
\frame{$\zeta$} & ${\small \star}$ & $%
\begin{array}
[c]{cc}%
\zeta{\small =}\frac{1}{\sqrt{1-\delta^{2}}}\left\{  \pi-\arcsin\left[
\sqrt{1-\delta^{2}}\frac{\delta+\sqrt{\delta^{2}+z(z+2)}}{(1+z)}\right]
\right\}  ; & \frac{\pi}{2}<\zeta\sqrt{1-\delta^{2}}\\
\zeta{\small =}\frac{1}{\sqrt{1-\delta^{2}}}\arcsin\left[  \sqrt{1-\delta^{2}%
}\frac{\delta\pm\sqrt{\delta^{2}+z(z+2)}}{(1+z)}\right]  ; & -\frac{\pi}%
{2}<\zeta\sqrt{1-\delta^{2}}<\frac{\pi}{2}\\
\zeta{\small =}\frac{1}{\sqrt{1-\delta^{2}}}\left\{  -\pi-\arcsin\left[
\sqrt{1-\delta^{2}}\frac{\delta-\sqrt{\delta^{2}+z(z+2)}}{(1+z)}\right]
\right\}  ; & \zeta\sqrt{1-\delta^{2}}<-\frac{\pi}{2}%
\end{array}
$ & ${\small \zeta=}\frac{\arccos\delta-\arccos\delta(\mathbf{t}_{0})}%
{\sqrt{1-\delta^{2}(\mathbf{t}_{0})}}$\\\hline
\frame{$z$} & ${\small z=}\frac{1}{\cos\left(  \sqrt{1-\delta^{2}}%
\zeta\right)  +\frac{\delta}{\sqrt{1-\delta^{2}}}\sin\left(  \sqrt
{1-\delta^{2}}\zeta\right)  }{\small -1}$ & ${\small \star}$ & ${\small z=}%
\sqrt{\frac{1-\delta^{2}(t_{0})}{1-\delta^{2}}}{\small -1}$\\\hline
\frame{$\delta$} & $\delta=\cos\left[  \arccos\delta(\mathbf{t}_{0}%
)+\sqrt{1-\delta^{2}(\mathbf{t}_{0})}\zeta\right]  $ & ${\small \delta=\pm
}\frac{\sqrt{\delta^{2}(t_{0})+z(z+2)}}{(1+z)}$ & ${\small \star}%
$\\\hline\hline
\end{tabular}
\caption{Connecting formulas between the six kinematical dimensionless variables, for the k = -1 case, including those not explicitly analyzed in the current section.}\label{TableKey:table1}%
\end{sidewaystable}%

\subsection{\label{sect:age_of_universe}The age of the Universe and the
horizon problem}

Two important issues which need to be addressed by any cosmological theory are
the age of the Universe and the existence of particle or event
\textquotedblleft horizons,\textquotedblright\ which might limit our
\textquotedblleft view\textquotedblright\ of light signals and events from the
past. We can show that these two topics do not cause any problem in our
conformal cosmology.

These issues were originally addressed by Mannheim (\cite{Mannheim:1996rv},
\cite{Mannheim:1991cz}, \cite{Mannheim:1996cd}, \cite{Mannheim:1996wf}) by
showing that the Conformal Gravity theory does not possess any horizon or
flatness problem, and does not contradict current estimates of the age of the
Universe. For completeness, we present in this section a similar analysis,
based on our particular solutions for the cosmic scale factor.

The age of the Universe is determined experimentally from various
observations, such as the age of chemical elements, which leads to age
determinations of terrestrial rocks and meteorites thus determining the age of
the solar system. Other sources of information include the age of the oldest
star clusters, white dwarfs, etc. Without entering into the details of these
determinations, they all give estimates or set lower limits for the age or the
Universe of about $10\ Gyr\lesssim t_{0}\lesssim15\ Gyr$, assuming that our
current time $t_{0\text{ }}$is measured from an initial singularity. These
estimates essentially agree with the scientific consensus based on standard
cosmology, which evaluates this age to be about $13.7$ billion years
\cite{Spergel:2003cb}.

In Sect. \ref{sect:time_dependent_form} we have already remarked that the
local time variable $t$, which is used for experimental age determinations,
cannot be identified with the cosmic standard time $\mathbf{t}$, since the
former is essentially the \textit{conformal time} of the latter, in the sense
of Eqs. (\ref{eqn3.15.3}) and (\ref{eqn3.15.4}). As described graphically in
Fig. \ref{fig5} and Fig. \ref{fig7}, the cosmic evolution of the scale factor
shows initial and final singularities when using the cosmic time $\mathbf{t}$,
similar to the case of the evolution described in terms of the universal
parameter $\delta$, but does not show any initial or final singularity when
using the other time variable $t$. This is due to the stretching of our local
time coordinate $t$, done by the time conformal transformation just mentioned,
so that the age of the Universe appears to be infinite in this temporal
coordinate. Therefore current experimental determinations, also done using
time $t$, will never contradict this infinitely lasting Universe.

On the contrary, the Universe appears to be limited temporally if a
\textquotedblleft cosmic\textquotedblright\ time $\mathbf{t}$ or $\delta$ is
used. We have already described the limits of these two variables; the delta
parameter varies in the range $-1<\delta<1$, while the limits for the
$\mathbf{t}$ variable are more easily expressed by the corresponding limits
for $\zeta=c(\mathbf{t}_{0}-\mathbf{t)}/\mathbf{R(t}_{0}\mathbf{)}$, resulting
in $\frac{1}{\sqrt{1-\delta^{2}}}\arctan\left(  -\frac{\sqrt{1-\delta^{2}}%
}{\delta}\right)  <\zeta<\frac{1}{\sqrt{1-\delta^{2}}}\left[  \arctan\left(
-\frac{\sqrt{1-\delta^{2}}}{\delta}\right)  +\pi\right]  $ (in this last
expression $\delta=\delta(t_{0})$ is our current value of this parameter). We
will estimate later \cite{Varieschi:2008va} the values of all our cosmological
parameters, but we can anticipate that the experimental range of the age of
the Universe, $t_{0}\approx10\ Gyr-15\ Gyr$ mentioned above, will translate in
a value of $\delta\cong-1$, obtained using Eq. (\ref{eqn4.38}) after
transforming the \textquotedblleft age\textquotedblright\ $t_{0}$ into a
corresponding value $\chi_{0}$. Therefore, the experimental observations seem
to point to cosmological times very close to the initial singularity, although
they cannot be used directly to measure the age of the Universe.

Another important topic is the analysis of possible horizons which might limit
our perception of past events and light signals. The so-called horizon problem
was one of the main reasons why an inflationary phase of the Universe was
proposed in the standard model. A particle horizon is usually referred to the
comoving radius $\chi_{H}$:%

\begin{equation}
\chi_{H}=S_{\mathbf{k}}^{-1}(\mathbf{r}_{H})=\int_{\mathbf{0}}^{\mathbf{r}%
_{H}}\frac{d\mathbf{r}}{\sqrt{1-\mathbf{kr}^{2}}}=c\int_{\mathbf{t}_{H}%
}^{\mathbf{t}_{0}}\frac{d\mathbf{t}}{\mathbf{R(t)}}, \label{eqn4.44}%
\end{equation}
assuming the integral in cosmic time $\mathbf{t}$ converges for $\mathbf{t}%
_{H}\mathbf{\rightarrow0}$ in models with an initial singularity, or for
$\mathbf{t}_{H}\mathbf{\rightarrow-\infty}$ in models without initial
singularity, thus yielding a finite value for $\chi_{H}$. On the contrary, if
the integral on the right-hand side of Eq. (\ref{eqn4.44}) is diverging for
the same limits for the variable $\mathbf{t}_{H}$, the horizon problem
disappears altogether.

In our model, this integral diverges to $+\infty$ when $\mathbf{t}_{H}$
approaches its lower bound, given by the corresponding (upper) limit for the
$\zeta$ variable mentioned above. This can be checked directly by computing
the integral, or simply by recalling that $\chi_{H}=c\sqrt{\left\vert
k\right\vert }(t_{0}-t_{H})$ in standard time $t$, following Eq.
(\ref{eqn4.24}), and that this variable is not bounded in the past or the
future, thus for $t_{H}\rightarrow-\infty$ we immediately get $\chi
_{H}\rightarrow\infty$ and no particle horizon is present. In this way all
regions in the Universe can be causally connected by light signals, including
the epoch when the Cosmic Microwave Background (CMB) was generated. We do not
need to invoke inflationary phases to justify the highly homogeneous nature of
the CMB.

For the same reason, no event horizons appear in our formulation. These are
associated with a similar integral:%

\begin{equation}
\chi_{EH}=S_{\mathbf{k}}^{-1}(\mathbf{r}_{EH})=\int_{\mathbf{0}}%
^{\mathbf{r}_{EH}}\frac{d\mathbf{r}}{\sqrt{1-\mathbf{kr}^{2}}}=c\int
_{\mathbf{t}}^{\mathbf{t}_{MAX}}\frac{d\mathbf{t}^{\prime}}{\mathbf{R(t}%
^{\prime}\mathbf{)}} \label{eqn4.45}%
\end{equation}
for an event which occurred at $(\mathbf{r}_{EH},\mathbf{t)}$ to be detected
at a later time through light signals, but before time $\mathbf{t}_{MAX}$
which can be be either infinity or the time of a full contraction of the
Universe. Again, the last integral simply gives $\chi_{EH}=c\sqrt{\left\vert
k\right\vert }(t_{MAX}-t)$ in the SSC coordinate, so that $\chi_{EH}%
\rightarrow\infty$ for $t_{MAX}\rightarrow+\infty$, and we can therefore
receive information from any event in the past if we wait a long enough time interval.

It is beyond the scope of this paper to investigate also the other reasons
which led to postulate the inflationary scenario, such as the flatness
problem, the fine tuning of parameters and others, but some of these issues do
not seem in any case to be significant in our kinematical cosmology, where
most of the physical parameters are varying with the cosmic time $\delta$ and
therefore do not require any particular explanation for the values they
currently hold. The cosmic time $\delta$, or any similar parameter, seems to
be driving the evolution of the Universe from its initial to its final value
and most of the other physical quantities just follow this evolution.

\section{\label{sect:kinematic_cosmology}Connection with Kinematic Cosmology}

We have already remarked in Sect. \ref{sect:new_red_shift} that a
\textit{kinematic cosmology}\ was introduced by L. Infeld and A. Schild (I-S
in the following) in 1945 (\cite{1945Nature}, \cite{1946PhDT.........6S},
\cite{PhysRev.68.250}, \cite{PhysRev.70.410}). These physicists were focusing
their attention on \textquotedblleft that part of relativistic cosmology which
deals with the metric form of our Universe, characterized by a
four-dimensional space-time manifold, and with the motion of free particles
and light rays,\textquotedblright\ \cite{PhysRev.68.250} thus dealing with the
kinematical description of cosmology while ignoring its dynamical aspect.

By assuming three fundamental postulates, the first one on light-geometry
(namely the existence of a cosmological coordinate system - CCS, conformal to
flat Minkowski space), plus the usual postulates of isotropy and homogeneity
of the Universe, Infeld and Schild introduced the following
metric:\footnote{We will use a different type of characters ($\mathfrak{t}%
$,$\mathfrak{r}$) to distinguish this new set of coordinates (CCS or CFS type)
from all the other coordinates used in this paper. The angular part of the
metric will remain unaffected by all the transformations in this section.}%

\begin{equation}
ds^{2}=\gamma(\mathfrak{t},\mathfrak{r})\left(  -c^{2}d\mathfrak{t}%
^{2}+d\mathfrak{r}^{2}+\mathfrak{r}^{2}d\psi^{2}\right)  , \label{eqn5.1}%
\end{equation}
where $\gamma(\mathfrak{t},\mathfrak{r})$ is a dimensionless conformal factor.
Conformally Flat Space-time (CFS) is the modern term used today to denote such
metrics, which were also studied in more recent works
(\cite{1994ApJ...434..397E}, \cite{1997ApJ...479...40E},
\cite{1998ApJ...508..129Q}, \cite{1993ApJ...405...51N},
\cite{1995MNRAS.277..627E}). In addition, Infeld and Schild were able to
restrict the possible $\gamma(\mathfrak{t},\mathfrak{r})$ functions satisfying
the three postulates, to just three fundamental classes corresponding to
standard closed, open and flat universes (in the notation of Ref.
\cite{PhysRev.68.250}, case I - $K>0$, case II - $K<0$ and case III - $K=0$, respectively).

Therefore, it is important to check our cosmological solutions against the
general classes proposed by Infeld and Schild, to make sure that they have the
general form required by the three fundamental postulates mentioned above. A
similar check should be performed for any other cosmological solution obtained
using other conformally-invariant theories in the literature.

We will consider here only case II - $K<0$, since this will correspond to our
preferred cosmological solution, for $\mathbf{k}=-1$, presented in Sect.
\ref{sect:evaluation_cosmic}. For this particular case the conformal factor
$\gamma$ must be of the form \cite{PhysRev.68.250}:%

\begin{equation}
\gamma(\mathfrak{t},\mathfrak{r})=f\left[  \frac{c\mathfrak{t}/2\alpha
}{1+\left(  c^{2}\mathfrak{t}^{2}-\mathfrak{r}^{2}\right)  /4\alpha^{2}%
}\right]  \left[  1+\frac{\left(  c^{2}\mathfrak{t}^{2}-\mathfrak{r}%
^{2}\right)  }{4\alpha^{2}}\right]  ^{-2}, \label{eqn5.2}%
\end{equation}
where $2\alpha$ is a convenient \textit{natural cosmological unit} (with
dimension of length) introduced by Infeld and Schild, so that the resulting
$\gamma$ factor will be dimensionless, for any choice of the arbitrary
function $f$ ($K=\pm1/\alpha^{2}$ in cases I and II of Ref.
\cite{PhysRev.68.250}).

The importance of the work by Infeld and Schild is also due to their original
derivation of the finite coordinate transformations from Conformally Flat
Space-time (CFS) to standard Robertson-Walker (RW) metric as described in our
Eq. (\ref{eqn3.15}) or (\ref{eqn3.15.1}), thus proving that RW space-time is
conformally flat, i.e., equivalent to flat Minkowski space-time up to a
conformal factor represented by $\gamma(\mathfrak{t},\mathfrak{r})$ in Eq.
(\ref{eqn5.1}). We will review this important coordinate transformation in the
following section.

\subsection{\label{sect:from_CFS_to_RW}From Conformally Flat Space-time to the
Robertson-Walker Metric}

The full transformation from CFS to RW can be found in the original 1945 paper
by Infeld-Schild \cite{PhysRev.68.250}, as well as in other references
(\cite{1994ApJ...434..397E}, \cite{1998ApJ...508..129Q},
\cite{1975STIA...7626675L}). Again, we will consider in the following this
transformation just for the particular $\mathbf{k}=-1$ case, which will be
compared to our cosmological solutions presented in Sect.
\ref{sect:evaluation_cosmic}. We start from the CFS metric in Eq.
(\ref{eqn5.1}) and, following Ref. \cite{PhysRev.68.250}, we apply a first
transformation simply to introduce dimensionless coordinates $\widetilde
{t},\widetilde{r}$:%

\begin{align}
\widetilde{t}  &  =\frac{c\mathfrak{t}}{2\alpha}\label{eqn5.3}\\
\widetilde{r}  &  =\frac{\mathfrak{r}}{2\alpha}\nonumber
\end{align}
using the \textit{natural cosmological unit} $2\alpha$. The metric then becomes%

\begin{equation}
ds^{2}=\widetilde{\gamma}(\widetilde{t},\widetilde{r})\left(  -d\widetilde
{t}^{2}+d\widetilde{r}^{2}+\widetilde{r}^{2}d\psi^{2}\right)  , \label{eqn5.4}%
\end{equation}
where $\widetilde{\gamma}(\widetilde{t},\widetilde{r})=4\alpha^{2}%
\gamma(\mathfrak{t},\mathfrak{r})$ has now the dimension of a length squared,
while the coordinates are all dimensionless. In this new units Eq.
(\ref{eqn5.2}) is rewritten as%

\begin{equation}
\widetilde{\gamma}(\widetilde{t},\widetilde{r})=\widetilde{f}\left[
\frac{\widetilde{t}}{1+\left(  \widetilde{t}^{2}-\widetilde{r}^{2}\right)
}\right]  \left[  1+\left(  \widetilde{t}^{2}-\widetilde{r}^{2}\right)
\right]  ^{-2} \label{eqn5.5}%
\end{equation}
where $\widetilde{f}=4\alpha^{2}f$ also acquires dimension of a squared
length. This is followed by a second transformation to another set of
dimensionless quantities:%

\begin{align}
X  &  =\widetilde{t}+\widetilde{r}\label{eqn5.6}\\
Y  &  =\widetilde{t}-\widetilde{r}\nonumber
\end{align}
which transforms the metric as follows,%

\begin{align}
ds^{2}  &  =\widetilde{\gamma}(X,Y)\left[  -dXdY+\frac{1}{4}(X-Y)^{2}d\psi
^{2}\right] \label{eqn5.7}\\
\widetilde{\gamma}(X,Y)  &  =\widetilde{f}\left[  \frac{X+Y}{2\left(
1+XY\right)  }\right]  \left(  1+XY\right)  ^{-2}.\nonumber
\end{align}

Two additional transformations are required to obtain the RW metric. The next
step is:%

\begin{align}
u  &  =\tanh^{-1}X\label{eqn5.8}\\
v  &  =\tanh^{-1}Y\nonumber
\end{align}
which yields a new form of the metric,%

\begin{equation}
ds^{2}=\frac{1}{4}\widetilde{f}\left[  \frac{1}{2}\tanh(u+v)\right]
\operatorname{sech}^{2}(u+v)\left[  -4dudv+\sinh^{2}(u-v)d\psi^{2}\right]  ,
\label{eqn5.9}%
\end{equation}
where $\operatorname{sech}(x)=1/\cosh(x)$, is the hyperbolic secant. The last
transformation is:%

\begin{align}
\eta &  =u+v\label{eqn5.10}\\
\chi &  =u-v\nonumber
\end{align}
which finally takes us to a RW metric, expressed in terms of the conformal
time $\eta$ and the comoving coordinate $\chi$, introduced in Eqs.
(\ref{eqn3.15.2}) - (\ref{eqn3.15.4}):%

\begin{align}
ds^{2}  &  =\mathbf{R}^{2}(\eta)\left(  -d\eta^{2}+d\chi^{2}+\sinh^{2}\chi
d\psi^{2}\right) \label{eqn5.11}\\
\mathbf{R}^{2}(\eta)  &  =\frac{1}{4}\widetilde{f}\left(  \frac{1}{2}\tanh
\eta\right)  \operatorname{sech}^{2}\eta.\nonumber
\end{align}

The previous equation also restricts possible functions for the cosmic scale
factor $\mathbf{R}^{2}(\eta)$, to be of the form specified above, just leaving
the function $\widetilde{f}$ totally arbitrary. One further step can bring the
metric of Eq. (\ref{eqn5.11}) into the standard Robertson-Walker metric of Eq.
(\ref{eqn3.15}) expressed in terms of our variables $\mathbf{t}$ and
$\mathbf{r}$. We just need to apply the inverse of Eqs. (\ref{eqn3.15.2}) and
(\ref{eqn3.15.3}), namely:%

\begin{align}
\mathbf{t}  &  =\frac{1}{c}\int\mathbf{R}(\eta)d\eta\label{eqn5.12}\\
\mathbf{r}  &  =\sinh\chi\nonumber
\end{align}
and the previous metric will become the standard RW expression for the
$\mathbf{k}=-1$ case,%

\begin{equation}
ds^{2}=-c^{2}d\mathbf{t}^{2}+\mathbf{R}^{2}(\mathbf{t})\left[  \frac
{d\mathbf{r}^{2}}{1+\mathbf{r}^{2}}+\mathbf{r}^{2}d\psi^{2}\right]  .
\label{eqn5.13}%
\end{equation}

\subsection{\label{sect:comparison}Comparison with our cosmological solution}

In order to compare the Infeld-Schild version of kinematical cosmology
outlined in the previous section with our model, it is easier to use
cosmological equations expressed in terms of the variables $\eta$ and $\chi$.
We can compare the I-S expression of the cosmic scale factor $\mathbf{R}%
(\eta)$ given in Eq. (\ref{eqn5.11}) with our expression, for the $k<1$ case,
from Eq. (\ref{eqn4.36}). This equation can be written also as%

\begin{align}
\mathbf{R}(\eta)  &  =\frac{\mathbf{R}(\chi=0)}{\cosh\chi-\delta\sinh\chi
}=\frac{\mathbf{R}(\eta_{0})}{a\cosh\eta+b\sinh\eta}=\frac{\mathbf{R}(\eta
_{0})}{a+2b\frac{\tanh\eta}{2}}\operatorname{sech}\eta\label{eqn5.14}\\
a  &  =\cosh\eta_{0}-\delta\sinh\eta_{0}\ ;\ b=\delta\cosh\eta_{0}-\sinh
\eta_{0}\nonumber
\end{align}
where we used $\mathbf{R}$ instead of $R$, dividing both sides of Eq.
(\ref{eqn4.36}) by $\sqrt{\left\vert k\right\vert }$ and substituted the
comoving coordinate $\chi$ with the conformal time $\eta$, namely $\chi
=\eta_{0}-\eta$, in view of Eq. (\ref{eqn4.24}). In this way we notice that
our expression in the last equation is precisely of the type expected by the
I-S models, as in Eq. (\ref{eqn5.11}), and we can uniquely determine the
function $\widetilde{f}$ as follows,%

\begin{equation}
\widetilde{f}\left(  x\right)  =\frac{4\mathbf{R}^{2}(\eta_{0})}{\left(
a+2bx\right)  ^{2}}. \label{eqn5.15}%
\end{equation}

The quantities $a$ and $b$ are defined in Eq. (\ref{eqn5.14}) as a function of
the current conformal time $\eta_{0}$, depending on the arbitrary choice of
the zero for this variable. A better choice would be to measure both $\chi$
and $\eta$ variables from the point of maximum expansion of the Universe. This
was done explicitly in Sect. \ref{sect:other_plots} and led to Eq.
(\ref{eqn4.37}), which can be easily rewritten in terms of a new variable
$\widetilde{\eta}$, also measured from the position of maximum expansion and
defined as $\widetilde{\eta}=-\widetilde{\chi}=\chi_{\max}-\chi
=\operatorname{arctanh}\delta-\chi$. With this new choice of variable, Eq.
(\ref{eqn4.37}) would simply become $\mathbf{R}(\widetilde{\eta}%
)=\frac{\mathbf{R}(\eta_{0})}{\sqrt{1-\delta^{2}}}\operatorname{sech}%
\widetilde{\eta}$. This expression is consistent with our general form in Eq.
(\ref{eqn5.14}) for the particular case of $\eta_{0}=\operatorname{arctanh}%
\delta$, which leads to $a=\sqrt{1-\delta^{2}}$ and $b=0$. In this particular
case, the function $\widetilde{f}$ in Eq. (\ref{eqn5.15}) would become simply
a constant,%

\begin{equation}
\widetilde{f}\left(  x\right)  =\frac{4\mathbf{R}^{2}(\eta_{0})}{1-\delta^{2}%
}. \label{eqn5.15.1}%
\end{equation}

These results will also lead to a unique expression for the conformal factors
$\widetilde{\gamma}$ or $\gamma$, from Eqs. (\ref{eqn5.5}) and (\ref{eqn5.2}) respectively:%

\begin{align}
\widetilde{\gamma}(\widetilde{t},\widetilde{r})  &  =4\mathbf{R}%
^{2}(\widetilde{t}_{0})\left\{  a\left[  1+\left(  \widetilde{t}%
^{2}-\widetilde{r}^{2}\right)  \right]  +2b\widetilde{t}\right\}
^{-2}\label{eqn5.16}\\
\gamma(\mathfrak{t},\mathfrak{r})  &  =\frac{\widetilde{\gamma}(\widetilde
{t},\widetilde{r})}{4\alpha^{2}}=\frac{\mathbf{R}^{2}(\mathfrak{t}_{0}%
)}{\alpha^{2}}\left\{  a\left[  1+\frac{\left(  c^{2}\mathfrak{t}%
^{2}-\mathfrak{r}^{2}\right)  }{4\alpha^{2}}\right]  +b\frac{c\mathfrak{t}%
}{\alpha}\right\}  ^{-2},\nonumber
\end{align}
so that the original I-S metric of Eq. (\ref{eqn5.1}) is now completely
defined, except for the dimensionless ratio $\mathbf{R}^{2}(\mathit{t}%
_{0})/\alpha^{2}$. The previous expressions can be further simplified, by
measuring our space-time variables from the point of maximum expansion of the
Universe, so that $a=\sqrt{1-\delta^{2}}$ and $b=0$, as discussed above.

It is beyond the scope of this work to analyze in more details the
implications of the Infeld-Schild kinematical cosmology, once their conformal
factor is fixed in the form of our Eq. (\ref{eqn5.16}). The objective of this
section was simply to show that our model is fully consistent with the
original I-S kinematical cosmology, as shown in the previous discussion.

We remark here that the CFS metric is just another way to describe our
Universe and that the I-S coordinates $(\mathfrak{t},\mathfrak{r})$ are
different from the RW coordinates $(\mathbf{t},\mathbf{r})$, or our original
SSC coordinates $(t,r)$, although they are all connected by the set of
transformations outlined in Sect. \ref{sect:from_SSC_to_RW} and Sect.
\ref{sect:from_CFS_to_RW}.

The original goal of kinematical cosmology, as introduced by Infeld and
Schild, was to describe the Universe through the motion of its
\textit{fundamental particles} (the galaxies, or nebulae as they were called
in 1945), rather than have fixed, comoving galaxies in RW metric and describe
the evolution of the Universe through the cosmic scale factor $\mathbf{R}$. In
other words, Infeld and Schild traded the advantage of the RW description,
i.e., to have the fundamental particles at rest (comoving), for a metric
conformal to Minkowski flat space-time, thus equivalent to the space of
special relativity, where the speed of light is simply constant, light
propagates as in flat space, and standard physical laws, such as Maxwell's
equations, can be extended without modifications from Minkowski space to CFS.

This different way of describing the Universe implies a motion of the
fundamental particles, as observed in CFS. This motion was also studied in
detail by I-S. In particular, our cosmological solution presented in Eq.
(\ref{eqn5.16}) belongs to the I-S Case II (\cite{PhysRev.68.250}), which is
called a \textquotedblleft converging-diverging\textquotedblright\ model,
since the fundamental particles move radially in a way that brings them
together towards the space-time origin and then away from it. Again, we will
leave to future studies a more complete analysis of our cosmological solution,
in view of the Infeld-Schild theory.

\section{\label{sect:conclusions}Conclusions}

We presented in this work the mathematical foundations of a new kinematical
approach to conformal cosmology. This was based on the assumption that the
observed redshift is mainly due to a gravitational origin and that the
gravitational potential of conformal gravity might be the cause of the
observed \textquotedblleft expansion\textquotedblright\ of the Universe.

We have seen how the original Mannheim-Kazanas potential can support this
explanation and how the chain of transformations, from Static Standard
Coordinates to the Robertson-Walker metric, can lead to a unique expression of
the cosmic scale factor. In particular, the $\mathbf{k}=-1$ solution is the
only capable of producing the observed galactic redshift and describes the
evolution of the Universe in terms of an initial expansion phase (which can
still be traced back to an initial singularity) up to a point of maximum
expansion, then followed by a symmetrical contraction phase, towards a final singularity.

We have compared our solutions with those of current conformal cosmologies and
noted that our kinematic cosmology is recovered from conformal gravity with a
cosmological constant at late (or early) cosmological times. This implies that
during these cosmological epochs the cosmological and gravitational redshifts
might be in fact equivalent, as assumed in our kinematical approach to
conformal cosmology.

Our detailed analysis of the solutions has also shown the importance of using
dimensionless quantities and, in particular, of the cosmological variable
$\delta\equiv\gamma/2\sqrt{\left\vert k\right\vert }$ which effectively
combines together the two parameters $\gamma$ and $k$ (or $\kappa$) originally
introduced by Mannheim and Kazanas. We also introduced the hypothesis that
$\delta$ might actually represent a true cosmic time, in terms of which the
evolution of the Universe would be described by the very simple Eq.
(\ref{eqn4.29}). If this interpretation is correct, all the space-time
dimensionless variables $\alpha$, $r$, $\chi$, $\zeta$, $z$, $\delta$ are
linked together and also related to the current value $\delta(t_{0})$ of the
cosmic time, through the transformations outlined in Table
\ref{TableKey:table1}.

The current value $\delta(t_{0})$ should be small and positive, as indicated
by Mannheim's estimate of the parameter $\gamma_{0}$, but its actual value has
to be determined by more precise fitting of astrophysical data, such as the
luminosities of type Ia Supernovae or others. This will be the objective of a
second part of this project \cite{Varieschi:2008va}, where we will also try to
explain the anomalous local blueshift region, which is implied by our
cosmological solution.

If our model will be successful in explaining current experimental data,
\textit{kinematical conformal cosmology} might become a viable alternative
model for the description of the Universe, with the advantage of avoiding
altogether the introduction\ of dark matter and dark energy, or other
controversial features of the current standard model.

\begin{acknowledgments}
This work was supported by a grant from Research Corporation. The author
wishes to acknowledge useful discussions with Dr. K. Knight and also
suggestions and clarifications by Dr. P. Mannheim. The author would like also
to thank the anonymous reviewers for helpful suggestions and comments.
\end{acknowledgments}

\bibliographystyle{apsrev}
\bibliography{CMB,CONFORMAL,CONSTANTS,COSMOBOOKS,HSTKEY,MOND,OTHERCMB,PIONEER,PK,SDSS,SUPERNOVAE}

\end{document}